\documentclass{article}
\usepackage{natbib,epsfig,emulateapj,apjfonts}

\slugcomment{Submitted to ApJ}
\begin{document}
%\draft

%\submitted{To appear in The Astrophysical Journal}

\title{Propagation of thermonuclear flames on rapidly rotating neutron
stars: extreme weather during type I X-ray bursts}

\author{Anatoly Spitkovsky$^1$, Yuri Levin$^1$, and Greg Ushomirsky$^2$}

\affil{$^{(1)}$Astronomy Department and Theoretical Astrophysics
Center,  601 Campbell Hall, \\ University of California, Berkeley,
CA 94720\\
anatoly@astron.berkeley.edu, yurlev@astron.berkeley.edu}
\affil{$^{(2)}$Theoretical Astrophysics, M/C 130-33, California Institute of
Technology, Pasadena, CA 91125\\
gregus@tapir.caltech.edu}
\date{printed \today}

\begin{abstract}
We analyze the global hydrodynamic flow in the ocean of an accreting,
rapidly rotating, non-magnetic neutron star in a low-mass X-ray binary
during a type I X-ray burst.  We use both analytical arguments and
numerical simulations of simplified models for ocean burning.  Our
analysis extends previous work by taking into account the rapid
rotation of the star and the lift-up of the burning ocean during the
burst. We find a new regime for the spreading of a nuclear burning 
front, where the flame is carried along a coherent shear flow across
the front.  If turbulent viscosity is weak, the speed of flame
propagation is $v_{\rm flame}\sim (gh)^{1/2}/ft_n\sim20$~km~s$^{-1}$,
where $h$ is the scale height of the burning ocean, $g$ is the local
gravitational acceleration, $t_n$ is the timescale for fast nuclear
burning during the burst, and $f$ is the Coriolis parameter, i.e.,
twice the local vertical component of the spin vector.  If turbulent
viscosity is dynamically important, the flame speed increases, and
reaches the maximum value, $v^{\rm max}_{\rm flame}\sim(gh/f
t_n)^{1/2}\sim 300$~km~s$^{-1}$, when the eddy overturn
frequency is comparable to the Coriolis parameter $f$.

We show that, due to rotationally reduced gravity, the thermonuclear
runaway which ignites the ocean is likely to begin on the equator. The
equatorial belt is ignited at the beginning of the burst, and the
flame then propagates from the equator to the poles.  Inhomogeneous
cooling (equator first, poles second) of the hot ashes drives strong
zonal currents which may be unstable to the formation of Jupiter-type
vortices; we conjecture that these vortices are responsible for
coherent modulation of X-ray flux in the tails of some bursts. We
consider the effect of strong zonal currents on the frequency of
modulation of the X-ray flux and show that the large values of the
frequency drifts observed in some bursts can be accounted for within
our model combined with the model of homogeneous radial expansion.
Additionally, if vortices or other inhomogeneities are trapped in the
forward zonal flows around the propagating burning front, fast chirps
with large frequency ranges ($\sim25-500$~Hz) may be detectable during
the burst rise.  Finally, we argue that an MHD dynamo within the
burning front can generate a small-scale magnetic field, which may
enforce vertically rigid flow in the front's wake and can explain
the coherence of oscillations in the burst tail.

\end{abstract}

\keywords{accretion, accretion disks -- instabilities -- hydrodynamics --
nuclear reactions, nucleosynthesis, abundances -- stars: neutron --
stars: rotation -- X-rays: bursts}

\section{Introduction}
\label{sec:introduction}

Accreting neutron stars (NSs) in low-mass X-ray binaries (LMXBs)
undergo type I X-ray bursts due to thermonuclear runaways in pure
helium or mixed hydrogen/helium layers (Hansen and Van Horn 1975,
Maraschi and Cavaliere 1976; Woosley and Taam 1996; see Lewin, van
Paradijs, and Taam~1995 and Bildsten 1998 for reviews).  Spherically
symmetric models of such runaways successfully explain general
features of type I X-ray bursts, such as burst fluences
($\sim10^{39}$~erg), timescales for accretion between bursts ($\sim$
few hours), and burst durations ($\sim 10$ seconds).   However,
spherically symmetric models cannot account for the lateral spreading
of thermonuclear flames and its interplay with NS rotation, and
recent observations have brought this issue into focus. 

Since the timescale for accretion between bursts is much longer than
the burst duration, it is unlikely that identical conditions exist
over the whole stellar surface for the burning instability to start
simultaneously (Shara 1981).  Therefore, burning should start
locally at some point, creating a brightness asymmetry, and spread over the entire surface of the star.
As nuclear burning spreads around the star, one would then expect to see a
rotational modulation of the X-ray flux, with the frequency of the modulation
equal that of the neutron-star spin. Such highly coherent ($Q\sim$~few
thousand) ``burst oscillations'' have indeed been observed with RXTE in
nine different bursters for many  bursts with 
 rise times of $<1$sec ~(see van der Klis 2000 for a review and
Muno et.~al.~2001 for the most recent tally). The inferred neutron-star
spins $\nu_s$ are between $250$ and $650$~Hz.

The discovery of burst oscillations, while confirming the basic
expectations of millisecond spins of accreting NSs and asymmetry of
nuclear burning, has brought about a host of new questions. Firstly,
burst oscillations are seen only from some of the $\sim50$ known
Galactic LMXBs, and only some bursts from the same source show
oscillations. For NSs with $\nu_s\sim600$~Hz, oscillations are seen
only during strong bursts with photospheric radius expansion, while for
$\nu_s\sim300$~Hz, oscillations are equally as likely to be seen
during weak or strong bursts (Muno et.~al.~2001).  Secondly,
oscillations are most commonly seen during {\it tails\/} of bursts,
when, presumably, the entire accreted fuel layer has been burned and
the obvious asymmetry is no longer present.  Finally, the frequency of
burst oscillations drifts upward by $\Delta\nu\sim$~several Hz during
the burst.

A simple model for the drift of the burst oscillation frequency has
been proposed by Strohmayer~et al.~(1997), and recently considered
 quantitatively by Cumming and Bildsten~(2000) and Cumming et.~al.~(2001).
 Since the
vertical sound crossing time through the burning layer (microseconds)
is much smaller than the nuclear burning time ($>$ fraction of a second) on
which the burst evolves, the outer layers of the NS are always in
hydrostatic balance. When the burning layer is hot, it expands
hydrostatically by $\Delta z\sim$tens of meters while conserving
angular momentum, and, hence, lags behind the neutron star by
$\Delta\nu/\nu_s\approx 2\Delta z/R$.  During the tail of the burst, a
postulated temperature inhomogeneity gives rise to oscillations, and,
as the layer cools down and contracts, we observe an upward frequency
drift by $\Delta\nu$ of a few Hz, roughly consistent with observations.
However, recent works by Van Straaten et al.~(2001), Galloway et al.~(2001),
Wijnands et al.~(2001), and Cumming et al.~(2001) suggest that purely
radial hydrostatic expansion may not be sufficient to explain rather
large $\Delta\nu$'s observed in some bursts; see discussion in \S~\ref{coolflow}. 

It has long been expected that NSs in LMXBs are progenitors of
millisecond radio pulsars (see Bhattacharya 1995 for a
review). Detection of burst oscillations has certainly bolstered the
idea that accreting NSs can reach periods similar to those seen in
millisecond radio pulsars. However, except for one source, there is no
convincing evidence of millisecond pulsar-like $\sim10^9$~G magnetic
fields in LMXBs.  Only SAX~J1808.4-365 (Wijnands and van der Klis
1998; Chakrabarty and Morgan 1998) shows coherent $\nu_s=401$~Hz
pulsations in persistent emission, which is consistent with the fact that a
$10^8-10^9$~G magnetic field channels accretion onto magnetic polar caps and
creates a permanent brightness asymmetry on the NS (Psaltis and
Chakrabarty~1999; though, since persistent flux modulation is only a
few percent, a weaker field could perhaps be sufficient). None of the other
LMXBs show such pulsations in persistent emission, implying that, if
they do have magnetic fields, they must be weaker than at least
$10^9$~G. Moreover, the presence of frequency drifts during bursts may
place even more stringent constraints on the magnetic field. If the
radial lift-up model described above were correct, it would imply that the
ocean and atmosphere of the neutron star can make up to $t_{\rm
burst}\Delta\nu\sim 10$ revolutions around its interior.  The magnetic
field, if present in the shearing layer, will be amplified by the
shear. A simple estimate (Cumming and Bildsten 2000) shows that even a
$\approx10^6$~G magnetic field might be dynamically important in this
situation. Throughout this work, we neglect the dynamical effects of 
the neutron-star magnetic field, and only briefly discuss possible 
MHD effects at the end.

In this paper, we abandon the requirement of spherical symmetry, and
consider instead the two-dimensional spreading of the nuclear fire
around the NS surface.  We study hydrodynamic flows that arise in the
burning ocean due to the combination of its inhomogeneous lift-up and
the rotation of the star. We analyze how these flows affect the
spreading of fire around the NS surface.  First we consider the local
conditions around the flame front at the interface between the hot,
burned ashes and cold, unburned fuel, exposing the non-trivial effects
of rotation and viscosity on the propagation of the flame front. Armed
with the understanding of local conditions, we then construct a global
scenario for X-ray bursts.

The burning front during type I X-ray bursts can propagate either by
deflagration (Fryxell and Woosley 1982b, Hanawa and Fujimoto 1984, Bildsten
1995) or by detonation (Fryxell and Woosley 1982a, Zingale
et.~al.~2001). Detonation is possible when the timescale for nuclear burning is
less than the vertical sound crossing time. This requires a large column of
accumulated fuel ($\sim 100$ meters) before the runaway starts, and, hence,
rather low accretion rates of $\lesssim 10^{-11.5} M_{\odot}$~yr$^{-1}$, i.e., $1-2$
orders of magnitude less than the observed accretion rates in most
bursters. However, direct numerical simulations of the type performed by
Fryxell and Woosley (1982a), and especially by Zingale et.~al.~(2001), if
extended into the parameter regime relevant for most of the observed
sources (i.e., higher accretion rate, and, hence, deflagration rather
than detonation), are the only definitive way to model X-ray bursts.

In most bursts, the helium/hydrogen ocean burns by spreading of a 
deflagration front; its propagation speed is set by the rate of heat transport
across the front.  Fryxell and Woosley (1982b) argue that in bursts with quick
($<1$~s) rise, which are of most interest to us, heat is conducted
sideways by convection.  They give three different phenomenological estimates
of the speed of the front [Eqs.~(3), (4), and the bottom line of the
left column on page $333$ of their paper].
%\begin{enumerate}
\begin{list}{}{
\setlength{\leftmargin}{1em}
\setlength{\itemindent}{-1em}
\setlength{\listparindent}{1em}}
\item{1.} The width of the front is equal to the lengthscale of a
convective roll, taken to be approximately the scaleheight $h$. The
front speed is then
\begin{equation}
v_{\rm flame}\sim \frac{h}{t_n}=
10^4 {\rm~cm~s}^{-1} \left(\frac{h}{10^3{\rm~cm}}\right)
\left(\frac{0.1{\rm~s}}{t_n}\right),
\label{frontspeed1}
\end{equation}
where $t_n$ is the timescale of nuclear burning during the burst. This estimate
is based on the earlier work of Ruderman (1981).

\item{2.} Heat is transported by turbulent convective diffusion, with a
kinetic diffusion coefficient $D=hv_c$, where $v_c$ is the
characteristic convective speed. The front speed is then
\begin{eqnarray}
v_{\rm flame}&\sim&\sqrt{D/t_n}
=2\times10^5{\rm~cm~s}^{-1} \\ \nonumber
&\times&
\left[\left(\frac{h}{10^3{\rm~cm}}\right)
\left(\frac{v_c}{5\times10^6{\rm~cm~s}^{-1}}\right)
\left(\frac{0.1{\rm~s}}{t_n}\right)\right]^{1/2}.
\end{eqnarray}

\item{3.} The turbulence scale is much larger than the front width, in which
case the burning front becomes wrinkled, accelerating heat transport. In this
case (Williams 1965)
\begin{equation}
v_{\rm flame}\sim v_c\sim 5\times 10^6{\rm~cm~s}^{-1}.
\label{frontspeed3}
\end{equation}
%\end{enumerate}
\end{list}

Cases $1$ and $2$, with typical expected $t_n\sim 0.1\hbox{sec}$,
$h\sim 10^3\hbox{cm} $, and $v_c\sim 5\times 10^6$~cm~s$^{-1}$, give
burst rise times longer than the ones observed.  Case $3$ is, Fryxel
and Woosley (1982a) argue, the upper limit for the front velocity, but
it seems that it is the only one of the three estimates which can
account for the observed short rise times of bursts.

In this paper we present a new mode of heat transport across the
front which does not depend on a large turbulence scale and on
wrinkling of the front, and can also explain the observed short rise
times of type I X-ray bursts.  The hot, burned ashes have a larger
scale height than the cold fuel. This vertical lift-up of the ocean as
the flame propagates around the star leads to a horizontal pressure
gradient.  Previous work (Strohmayer et al 1997, Cumming and Bildsten
2000) only considered the situation where the entire ocean is burned
and has lifted up, and, hence, the horizontal pressure balance is
already restored.  We argue that, as the thermonuclear flame  propagates around
the star, the lift-up of the ocean behind the front drives a
differential shear across the front.  This shear 
transports entropy from the hot ashes to the fuel, and, hence, 
propagates the front. The speed of the flame depends on the strength
of the frictional coupling (e.g., due to convection) between the
different layers (from top to bottom) of the burning ocean.

In Section \ref{secanalyt} we estimate 
analytically that 
%\begin{enumerate}
\begin{list}{}{
\setlength{\leftmargin}{1em}
\setlength{\itemindent}{-1em}
\setlength{\listparindent}{1em}}
\item{(a)} In the case of the weak frictional  top-bottom  coupling,
the width of the front $\Delta$ is given by the Rossby adjustment radius $a_R$:
\begin{eqnarray}
\Delta&\sim& a_R=\sqrt{gh_{\rm hot}}/f = 2 {\rm~km} 
\\ \nonumber
&\times& \left[\left({g\over 2\times 10^{14}\hbox{~cm}
\hbox{~s}^{-2}}\right)
\left({h_{\rm hot}\over 10^3\hbox{~cm}}\right)\right]^{1/2}
 \left({2000\hbox{~rad~s}^{-1}
\over f}\right),
\label{width1}
\end{eqnarray}
where $g$ is the gravitational acceleration at the NS surface, $h_{\rm
hot}$ is the scaleheight of the burned ocean, and
$f=2\Omega\cos\theta$ is the local Coriolis parameter ($\Omega$ is the
angular frequency of the NS and $\pi/2-\theta$ is the latitude). 
The speed of the front is 
\begin{eqnarray}
v_{\rm flame}&\sim& \frac{\Delta}{t_n}={\sqrt{gh_{\rm hot}}\over ft_n} \\ 
\nonumber
&=& 2\times 10^6{\rm~cm~s}^{-1} \left({\Delta\over
2\hbox{~km}}\right)\left({0.1\hbox{~s}\over t_n}\right),
\label{frontspeed4}
\end{eqnarray}
which gives a rise time of $\sim 0.7$~s for burning to spread from the
equator to the poles [cf. Eq.~(\ref{vfront12})].

\item{(b)} Let $t_{\rm fr}$ be the timescale of 
frictional coupling between the top and the bottom of the ocean;
$t_{\rm fr}\sim h_{\rm hot}/v_c$ if the layers of the ocean are
dynamically coupled by convection. If the coupling is strong, i.e., 
$t_{\rm fr}\lesssim t_n$, friction modifies
the structure of the burning front and its propagation speed. Maximum
speed is reached when $t_{\rm fr}\approx 1/f$, and is given by
\begin{eqnarray}
v_{\rm flame}^{\rm max}&\sim&\left(\frac{g h_{\rm hot}}{f t_n}
\right)^{1/2}
=3\times10^7{\rm~cm~s}^{-1} \\ \nonumber
&\times&\left[
\left({h_{\rm hot}\over 10^3\hbox{cm}}\right)
\left({2000\hbox{~rad~s}^{-1}\over f}\right)
\left({0.1\hbox{~s}\over t_n}\right)\right]^{1/2}
\label{frontspeed5}
\end{eqnarray}
[Equation~(\ref{vflame21}) gives  the expression for the burning
front speed when the frictional 
coupling strength is arbitrary].  The burst rise time is $\sim0.1$~s
in this case.
\end{list}
%\end{enumerate}

In Section \ref{numerical} we set up and solve numerically a $2$-layer
shallow-water model which contains the essential physics of how a
stably stratified ocean responds to inhomogeneously applied
heating. We use this $2$-layer model to simulate ignition and
propagation of the deflagration front, both for weak and strong
frictional coupling. Our simulations are in agreement with the analytical
estimates of the front speed in Section \ref{secanalyt}. The shallow-water
model also
provides us with the theoretical estimate of the size of the initial
ignition spot on the neutron-star surface.  However, we are
unable to determine the strength of friction from the first
principles, and we leave it as a free parameter of our
model. Hopefully, future direct
numerical simulations will address this issue.

Armed with our understanding of flame propagation, in
\S~\ref{sectionfour} we set out to construct a global scenario for
X-ray bursts. We argue that, due to rotationally reduced gravity,
the thermonuclear runaway is likely to begin in the equatorial region. We then
show that, due to the reduced Coriolis parameter at the equator, the
flame propagates faster along the equator than away from the equator.
Thus, a possibly inhomogeneous equatorial belt is ignited at the
beginning of the burst rise, and the flame proceeds to burn from the
equator to the poles.  Inside the burning front there are strong zonal
currents going forward relative to the star's rotation, and there are
weaker backward zonal flows in the cooling wake of the front.  A burning
inhomogeneity observed as a flux modulation during the burst will be
trapped in the backward zonal flows, thus the observed modulation
frequency is smaller than the neutron-star spin. As the ocean cools,
the zonal flow slows down, and the modulation frequency asymptotes to
the neutron-star spin frequency. This scenario for the burst frequency
drift is an extension of the radial lift-up scenario proposed by
Strohmayer et.~al.~(1997) as modeled in detail by Cumming and Bildsten
(2000).

We speculate that differential zonal currents can be unstable to the
formation of vortices of the type observed in the atmospheres of giant
planets. These vortices may be the cause of oscillations in the X-ray flux
in the tail of the burst.  In addition, if there are vortices trapped
in the strong forward zonal currents in the burning front, we expect
to see a chirp with a large frequency span ($\sim 25$ -- $500$ Hz)
during the rise of the burst.  Finally, we also suggest that the
burning front may generate, via an MHD dynamo, a magnetic field $B\sim
10^9$~G, with the ocean scaleheight as the characteristic coherence
length.  This magnetic field  can quench
the vertical shear in the backward zonal flow, and thus may be
responsible for the observed coherence of burst oscillations in the
burst tail.

\section{Propagation of the flame: analytical  arguments }\label{secanalyt}
\subsection{Vertical force balance}
\label{sec:vertical-balance}
During the burst, the burning material reaches temperatures of
$\sim 2\times 10^9$K, degeneracy in the helium layer is lifted,
and the burning ocean expands by $10-40$m --- many scaleheights of the
cold pre-burst ocean (Joss 1977, see Cumming and Bildsten 2000
for the most recent calculations).  This vertical expansion is very
subsonic.  However, since only a part of the star is buring at a given
time, the horizontal pressure imbalance leads to non-trivial
hydrodynamic flows, as we now discuss. 

\begin{figure}
\centerline{ \psfig{file=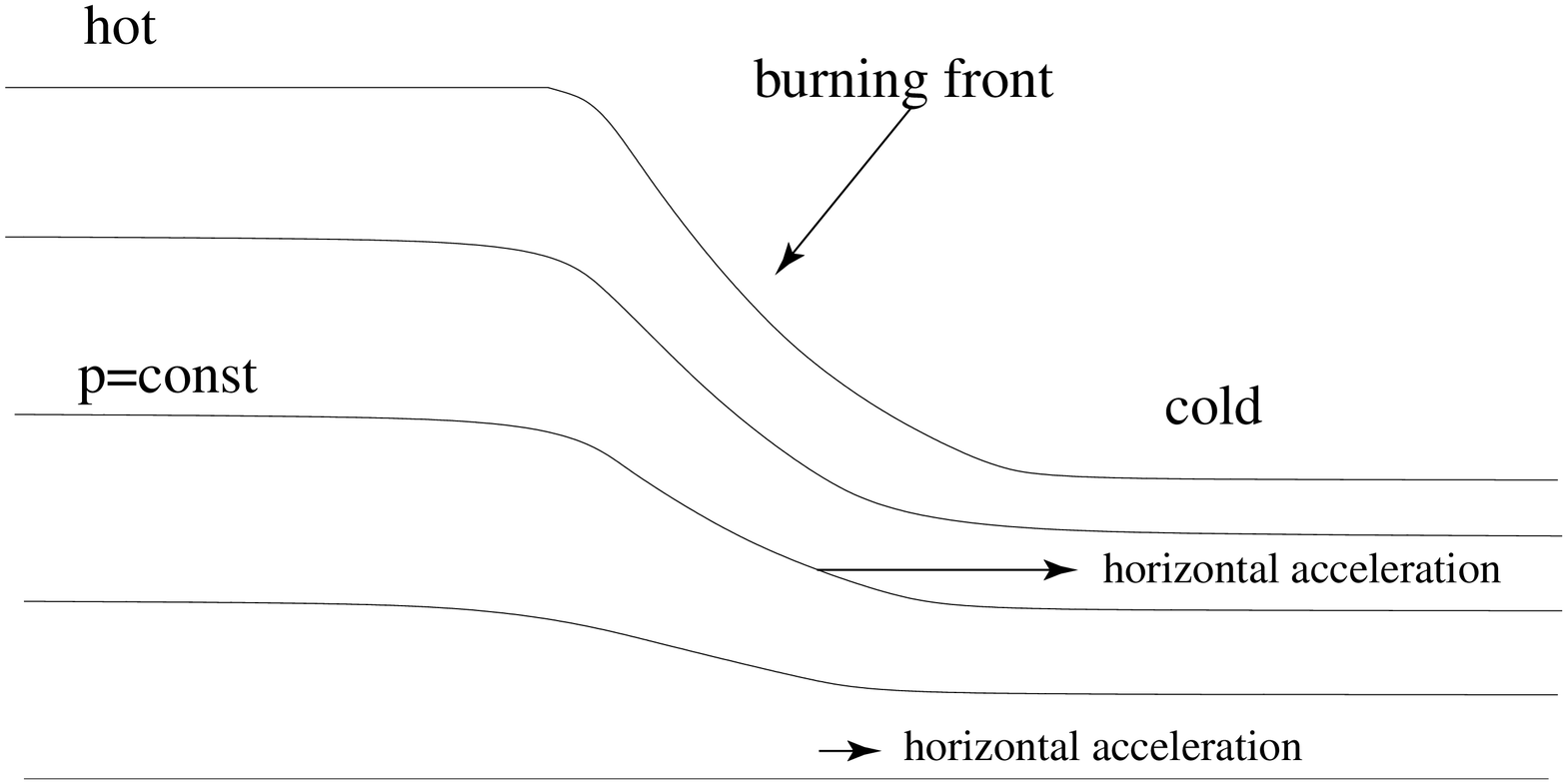,width=4in}}
\caption{ 
Illustration of a burning front moving from the hot to the cold region in the atmosphere/ocean}
\label{figfront}
\end{figure}

Consider a propagating burning front, as illustrated in
Fig.~\ref{figfront}, where the ocean behind the burning front is
already hot, while the ocean ahead of the front is still cold.  Let
the ocean, for simplicity, reside on a plane, with $\vec{x}$ and $z$
being the horizontal and the vertical coordinates respectively; $z$ is
taken to increase upward.  Since the vertical sound crossing time is
much smaller than characteristic nuclear burning timescale during the
burst, $t_n$, vertical hydrostatic equilibrium is always a good
approximation (we also neglect the vertical component of the Coriolis
force, see \S~\ref{sec:horiz-balance-rossby}). Therefore,
\begin{equation}
\left({\partial p\over \partial z}\right)_{\vec{x}}=-g\rho,
\label{hydrostaticequilibrium}
\end{equation}
where $p(\vec{x}, z)$ is the pressure, and $\rho(\vec{x}, z)$ is the
density. From Eq.~(\ref{hydrostaticequilibrium}) we see that the
separation between constant pressure surfaces scales as $1/\rho$. The
hot ocean is less dense than the cold one, so the pressure surfaces
diverge as they traverse the front from the cold into the hotter part,
see Fig.~\ref{figfront}. This implies that the horizontal pressure
gradient $\partial p/\partial \vec{x}$ (indicated by horizontal arrows
in Fig.~\ref{figfront}) increases with height inside the burning
front.  This difference in the horizontal pressure gradient drives a
shear flow and circulation across the front. 

Let us quantify the above argument.
Let $\phi(\vec{x}, z)$
be the effective gravitational potential per unit mass, $d\phi=g dz$. 
The horizontal\footnote{A horizontal surface is a surface of constant
$\phi$; in plane-parallel geometry with constant $g$, setting $\phi$=const
is equivalent to $z$=const.}
acceleration of
a fluid element due
to the pressure gradient is given by
\begin{equation}
\vec{a}_{\rm pressure}=-{1\over \rho (\vec{x}, \phi)}
                        \left({\partial p(\vec{x}, \phi)\over
                         \partial\vec{x}}\right)_\phi.
\label{pressuregradient}
\end{equation}
Furthermore,
\begin{equation}
\left({\partial p\over \partial \vec{x}}\right)_\phi=
-\left({\partial p\over \partial \phi}\right)_{\vec{x}}
\left({\partial\phi\over\partial\vec{x}}\right)_p=
\rho(\vec{x}, \phi)\left({\partial\phi\over\partial\vec{x}}\right)_p.
\label{mathidentity}
\end{equation}
Therefore, from Eqs.~(\ref{pressuregradient}) and (\ref{mathidentity}),
we have
\begin{equation}
\vec{a}_{\rm pressure}=-\left({\partial\phi\over\partial\vec{x}}\right)_p.
\label{apressure}
\end{equation}
Eq.~(\ref{apressure}) is easy to understand: the height $\phi/g$ of
surfaces of constant pressure decreases from the hot to the cold part
of the ocean, so $\vec{a}_{\rm pressure}$ is directed from the hot
region to the cold region (arrows in Fig.~\ref{figfront}).
Differentiating Eq.~(\ref{apressure}) with respect to $\ln{p}$ at
constant $\vec{x}$, and using Eq.~(\ref{hydrostaticequilibrium}), we
get
\begin{eqnarray}
\left({\partial\vec{a}_{\rm pressure}\over \partial \ln{p}}\right)_{\vec{x}}&=&
-p\left({\partial\over \partial \vec{x}}\right)_p \left({\partial\phi
\over\partial p}\right)_{\vec{x}} \\ \nonumber &=&
\left[{\partial(p/\rho)\over \partial \vec{x}}\right]_p=
\left[{\partial(c_s^2/\gamma)\over \partial\vec{x}}\right]_p,
\label{apressure2}
\end{eqnarray}
where $c_s$ is the speed of sound and $\gamma$ is the adiabatic index
of the gas. As an example, for an ideal chemically homogeneous gas
\begin{equation}
{\partial \vec{a}_{\rm pressure}\over \partial \ln{p}}={R\over\mu}
\left({\partial T\over\partial\vec{x}}\right)_p,
\label{thermalwind}
\end{equation}
where $T$ is the temperature and $\mu$ is the mean molecular weight of
the gas.  The term on the right hand side of Eq.~(\ref{thermalwind})
is the forcing term for the vertical shear; it is determined by the
horizontal temperature gradient.

\subsection{Horizontal force balance and flame propagation in the ocean
without friction}
\label{sec:horiz-balance-rossby}

Let us now write down the horizontal momentum equation. 
In a frame rotating with the star, a fluid element moving with
velocity $\vec{v}$ experiences the Coriolis acceleration,
$\vec{a}_{\rm coriolis}=-2 \vec{\Omega}\times\vec{v}$, 
where 
%$\vec{f}=2\vec{\Omega}$ is the Coriolis vector, 
$\vec{\Omega}$ is the angular velocity of the star.
In general, the Coriolis vector $2 \vec{\Omega}$ 
has both vertical and horizontal components
%$\Omega_z$ and $\Omega_x$, 
in the local frame of the ocean. 
In what follows we shall neglect
the horizontal component of the Coriolis vector  [this is commonly
known as the 
``traditional approximation'' in geophysics (see, e.g.,
Pedlosky 1987)]. In doing so, we neglect two
terms in the momentum equation. First, in the equation for the vertical force
balance (\S~\ref{sec:vertical-balance}) we neglect the vertical
component of the Coriolis force due to horizontal motion, $\sim  \Omega
v$. As we shall see below, $v_{\rm max}\sim (gh)^{1/2}$, so this component of
the Coriolis force is negligible compared to gravity so long as
$\Omega \ll(g/h)^{1/2}=4\times10^5 (h/10^3$~cm$)^{-1/2}$, which is always
satisfied for neutron stars in LMXBs.  Second, we neglect the horizontal component of the
Coriolis force due to radial motion, $\sim \Omega v_z$, which is small
compared to the
horizontal Coriolis force due to horizontal motion, $\sim \Omega v$. 
This can be seen
as follows: during the burst, the ocean expands vertically
 on the nuclear timescale $t_n$, and the
vertical velocity is $v_z\sim h/t_n$. Therefore, the
horizontal Coriolis acceleration due to radial motion 
is much less then the horizontal Coriolis acceleration
due to horizontal motion 
so long as
$1/t_n\ll (g/h)^{1/2}$. 
This condition is always satisfied for type I X-ray bursts.

With the above approximations, and in the absence of viscosity
(this assumption is alleviated in
\S~\ref{sec:horiz-balance-convective}), the horizontal force balance
is 
\begin{equation}
\vec{a}_{\rm pressure}={d\vec{v}\over dt}+\vec{f}\times\vec{v},
\label{forcebalance1}
\end{equation}
where $\vec{v}$ is the horizontal component of the velocity and
$\vec{f}=2\Omega\cos{\theta}\hat{z}$ is the vertical Coriolis vector. 

Accreting neutron stars which display type I X-ray bursts are typically
spinning with frequencies $\nu_s$ of a few hundred hertz.  The characteristic
timescales of the burst light curves are orders of magnitude greater than the
rotation period of the neutron star: a typical burst rise time is 
$0.1-1$~seconds, and a typical burst cooling time is $5-100$ seconds. 
Theoretical
calculations (see Bildsten 1998 for a review) suggest that the nuclear burning
timescale during a typical burst is $t_n\sim 0.1$~sec, still much greater than
$1/\nu_s$. Therefore, during the burst the ocean flow must be in quasigeostrophic
equilibrium (Pedlosky 1987) everywhere except for the immediate vicinity of the
equator.  Quasigeostrophic equilibrium implies that inertial external forces
acting on a fluid element in the horizontal direction---in particular, due to
the horizontal pressure gradient---are almost exactly balanced by the
horizontal component of the Coriolis force.

One can see this by noting that the second term on the right-hand side
of Eq.~(\ref{forcebalance1}) is dominant. We can rewrite this equation as
\begin{equation}
\vec{v}=-{1\over f^2}\vec{f}\times\vec{a}_{\rm pressure}+{1\over f^2}\vec{f}\times
{d\vec{v}\over dt}.
\label{v}
\end{equation}
To zeroth order,
\begin{equation}
\vec{v}=\vec{v}_{\rm geostrophic}=-{1\over f^2}\vec{f}\times\vec{a}_{\rm pressure}.
\label{vg}
\end{equation}
Substituting this expression into the right-hand side of Eq.~(\ref{v}), we get the
first-order correction to the velocity:
\begin{eqnarray}
\vec{v}&=&\vec{v}_{\rm geostrophic}+\vec{v}_{\rm ageostrophic}= \nonumber \\
&-&{1\over f^2}\vec{f}\times\vec{a}_{\rm pressure}+{1\over f^2}{d\vec{a}_{\rm pressure}
\over dt}.
\label{v1}
\end{eqnarray}
The first term in Eq.~(\ref{v1}) represents the dominant geostrophic flow in the
direction along the front line and perpendicular to the pressure
gradient. 
The second term is the ageostrophic
component of the velocity, which is perpendicular to the front line. Differentiating
Eq.~(\ref{v1}) with respect to $\ln{p}$ and using Eq.~(\ref{apressure2})
we get
\begin{equation}
{\partial \vec{v}_{\rm geostrophic}\over \partial \ln{p}}=
-{1\over f^2}\vec{f}\times{\partial (p/\rho)\over\partial \vec{x}},
\label{thermalwind1}
\end{equation}
and 
\begin{equation}
{\partial \vec{v}_{\rm ageostrophic}\over \partial \ln{p}}={1\over f^2}
{\partial\over\partial\ln{p}}{d\vec{a}_{\rm pressure}\over dt}.
\label{thermalwind2}
\end{equation}

 Equation~(\ref{thermalwind1}) is commonly referred to as the thermal wind
relation in the meteorology/geophysics literature. It implies that the difference
in geostrophic velocity $\delta v_{\rm geostrophic}$ between the top and the bottom of 
the ocean somewhere
in the middle of the front is related to the difference in $p/\rho$ between the
hot and the cold parts of the ocean:
\begin{eqnarray}
\delta v_{\rm geostrophic} &=& v^{\rm top}_{\rm geostrophic}-
v^{\rm bottom}_{\rm geostrophic} \nonumber \\
&\sim& \frac{(p/\rho)_{\rm hot}-(p/\rho)_{\rm cold}}{f\Delta}\approx
    \frac{gh_{\rm hot}}{f\Delta},
\label{thermalwind3}
\end{eqnarray}
where $h_{\rm hot}$ is the scaleheight of the ocean behind the front
which has just undergone a thermonuclear runaway, $\Delta$ is the
width of the front, and we used the fact that $h_{\rm hot}\gg h_{\rm
cold}$.

Likewise, we can use Eq.~(\ref{thermalwind2}) to estimate the
characteristic ageostrophic shear $\delta v_{\rm
  ageostrophic}=\vec{v}^{\rm top}_{\rm ageostrophic} -\vec{v}^{\rm
  bottom}_{\rm ageostrophic}$ somewhere in the middle of the front.
Since the front passes through a given fluid element on the timescale
of order $t_n$, we replace $d/dt$ by $1/t_n$ in our order-of-magnitude
estimates. Then
\begin{equation}
{\partial\vec{v}_{\rm ageostrophic}\over\partial\ln{p}}=
\delta\vec{v}_{\rm ageostrophic}\sim
{1\over f^2t_n}{\partial\vec{a}_{\rm pressure}\over\partial\ln{p}}\sim
{gh_{\rm hot}\over f^2t_n \Delta}\hat{n}.
\label{shear1}
\end{equation}
Here $\hat{n}$ is the unit vector perpendicular to the front line.

The quantity $\delta\vec{v}_{\rm ageostrophic}$ is the characteristic
speed of the shear flow perpendicular to the burning front. How does
it relate to the flame propagation speed?  Generally, flames propagate
by transporting entropy from the hot, burned material to the cold,
unburned fuel, and their speed is set by the time it takes to
transport this heat across the width of the front.  In the case of
slow, laminar deflagration fronts (see, e.g., Bildsten 1995) heat is
transported by conduction, while for convective fronts discussed by
Fryxell and Woosley (1982b), heat is advected by the turbulent motion
of the fluid and then mixed into the fuel.  In our case, however, it
is the shear flow, $\delta\vec{v}_{\rm ageostrophic}$, that moves the
hot material ahead of the front (this happens at the top of the
burning ocean), and pulls the cold fresh fuel into the burning front
(this happens at the bottom of the ocean).  

We assume that vertical
mixing occurs within the front, and that this mixing transports heat
between the hot burning fluid moving forward on the top and the cold
fluid on the bottom.  This mixing could be either due to convection,
or due to various shear instabilities.  Indeed, one-dimensional
numerical simulations, starting with Joss~(1977), show that X-ray
bursts are strongly convective, with convective overturn timescale of
$\sim 10^{-3}$~s. If convection is present within the front, it will
efficiently mix entropy in the vertical direction. However, while in
one dimension the hot burned fluid has no place to go but to mix
with the cold unburned fuel, in more than one dimension convection
may be quenched by the sideways propagation of the hot material on top
of the cold material\footnote{We thank Lars Bildsten and Andrew
Cumming for making this point.}.  In addition to the possibility of
convection, the geostrophic flow within the front possesses strong
vertical shear, with velocity difference of $\delta v\sim \sqrt
{gh_{\rm hot}}$ across a scaleheight [cf.~Eq.~(\ref{thermalwind1})
and~(\ref{gh})], which corresponds to the Richardson number of order
unity. The front with such strong shear may be unstable to the
Kelvin-Helmholtz instability.  Finally, the interface between the 
shearing layers is not exactly
horizontal, but has a slope of order $h_{\rm
hot}/\Delta\lesssim 10^{-2}$. Such flow may be unstable to the
baroclinic instability (Fujimoto 1988, Cumming and Bildsten 2000). The
nonlinear development of these or other instabilities might result in
efficient vertical mixing.  The presence of efficient vertical
mixing within the front is the most uncertain part of our model; it
must be addressed directly by future numerical simulations.  However,
given the large amount of shear in the burning front, such mixing is
not at all unreasonable.

We therefore believe that the vertical thermal mixing timescale
$t_{\rm mixing}$ within our geostrophic front is much smaller than
$t_n$ (e.g., the convective overturn timescale is $10^{-3}$~sec).  In
the foregoing, we assume that the entropy advected forward by the
ageostrophic flow is quickly mixed in the vertical direction. If the
thermal mixing timescale $t_{\rm mixing}$ within the front is
non-negligible, then the general argument laid out in this paper is
unchanged, except that all estimates should use $t_n+t_{\rm mixing}$
instead of just $t_n$.

If entropy mixing is efficient, then the
 front propagation speed is  the velocity with which entropy is
transported, i.e.,  the characteristic speed of the ageostrophic shear flow: 
\begin{equation}
v_{\rm flame}\sim\delta v_{\rm ageostrophic}\sim {gh_{\rm hot}\over f^2 t_n \Delta }.
\label{vfront11}
\end{equation}
The width of the propagating front is (e.g.,~Fryxell and Woosley 1982b)
\begin{equation}
\Delta\sim v_{\rm flame}t_n.
\label{width2}
\end{equation}
Substituting this into Eq.~(\ref{vfront11}), and solving for the front
speed and width, we get
\begin{eqnarray}
v_{\rm flame}&\sim& {(gh_{\rm hot})^{1/2}\over ft_n}\approx
20 {\rm~km~s}^{-1} \\ \nonumber
&\times&
\left(\frac{h_{\rm hot}}{10^3 {\rm~cm}}\right)^{1/2}
\left(\frac{2000{\rm~rad~s}^{-1}}{f}\right)\left(\frac{0.1{\rm~s}}{t_n}\right),
\label{vfront12}
\end{eqnarray}
and
\begin{equation}
\Delta\sim{(gh_{\rm hot})^{1/2}\over f}\approx 
2{\rm~km}\left(\frac{h_{\rm hot}}{10^3 {\rm~cm}}\right)^{1/2}
\left(\frac{2000{\rm~rad~s}^{-1}}{f}\right)
\label{width3}
\end{equation}
for $g=2\times 10^{14}\hbox{~cm~s}^{-2}$. From Eq.~(\ref{thermalwind3}) 
the characteristic geostrophic velocity inside the front is of order the
gravity wave speed in the hot material:
\begin{equation}
v_{\rm geostrophic}\approx (gh_{\rm{hot}})^{1/2}.
\label{gh}
\end{equation}
Note that the width of
the front is equal, to the order of magnitude, to the Rossby
adjustment radius (Pedlosky 1987).

\subsection{Horizontal force balance and flame propagation in the ocean
with
friction}
\label{sec:horiz-balance-convective}

In the previous subsection we have assumed that there are no viscous
forces acting on fluid elements, and hence that the top and the bottom
of the burning ocean are free to slip past each other. However,
there is likely to be some vertical mixing within the
front, due to turbulence driven by convection or by strong shear
(see the discussion in the previous subsection). This mixing
will exchange momentum between the top and the bottom of the ocean;
we must, therefore, consider what effect viscous-type  friction within the ocean
will have on the  propagation of the burning front.

  Let us return to the
horizontal force balance equation (\ref{forcebalance1}) and include
viscosity:
\begin{equation}
\vec{a}_{\rm pressure}={d\vec{v}\over dt}+\vec{f}\times\vec{v}-\vec{a}_{\rm viscous},
\label{forcebalance11}
\end{equation}
where we have added to the right-hand side the acceleration of the fluid element
due to viscous stress. We now consider the difference in $\vec{a}_{\rm
pressure}$ between the bottom and the top of the burning ocean, where we take
the top to be roughly one scaleheight above the bottom. From the previous
equation we have
\begin{eqnarray}
\vec{a}_{\rm pressure}^{\rm top}&-&\vec{a}_{\rm pressure}^{\rm bottom}=
{d\vec{v}_{\rm top}\over dt}-
{d\vec{v}_{\rm bottom}\over dt} \nonumber \\
&+&\vec{f}\times (\vec{v}_{\rm top}-\vec{v}_{\rm bottom})
-(\vec{a}_{\rm viscous}^{\rm top}-\vec{a}_{\rm viscous}^{\rm bottom}).
\label{forcebalance12}
\end{eqnarray}
We assume that effective viscosity dynamically couples the top and
the bottom of the burning ocean, and model this coupling by introducing a
linear
friction term:
\begin{equation}
\vec{a}_{\rm viscous}^{\rm top}-\vec{a}_{\rm viscous}^{\rm bottom}
=-\frac{1}{t_{\rm fr}} (\vec{v}_{\rm top}-\vec{v}_{\rm bottom}).
\label{friction}
\end{equation}
If friction  is mediated by turbulent eddies, then  $1/t_{\rm fr}=\beta_1
v_{\rm turb}/h$, where $v_{\rm turb}$ is the characteristic velocity of the
largest turbulent eddy,  and $\beta_1$ is a
number of order unity.  On the left-hand side of
Eq.~(\ref{forcebalance12}) we use Eq.~(\ref{apressure2}) to get
\begin{equation}
\vec{a}^{\rm top}_{\rm pressure}-\vec{a}^{\rm bottom}_{\rm pressure}\sim
{d\vec{a}_{\rm pressure}\over d\ln{p}}\sim\left[{\partial(p/\rho)\over\partial
\vec{x}}\right]_p\sim g{h_{\rm hot}\over  \Delta}\hat{n}.
\label{forcebalance13}
\end{equation}
Finally, as in \S~\ref{sec:horiz-balance-rossby}, we replace 
in (\ref{forcebalance12}) the time
derivative $d/dt$ by $\eta/t_n$, where $\eta$ is a number of order unity. 
 With the above simplifications, the
shear flow $\delta\vec{v}=\vec{v}_{\rm top}- \vec{v}_{\rm bottom}$
obeys
\begin{equation}
g{h_{\rm hot}\over \Delta}\hat{n}\sim  
\left[\vec{f}\times\delta\vec{v}
+\left({1\over t_{\rm fr}}+\frac{\eta}{t_n}\right)\delta\vec{v}\right].
\label{forcebalance14}
\end{equation}
Scalar multiplication of the above
equation by $\delta\vec{v}$ yields
\begin{equation}
g{h_{\rm hot}\over\Delta}(\delta\vec{v}\cdot \hat{n})\sim
\left({1\over t_{\rm fr}}+\frac{\eta}{t_n}\right)(\delta v)^2,
\label{vectoreq1}
\end{equation}
while application of 
Pythagorean theorem  to Eq.~(\ref{forcebalance14}) yields:
\begin{equation}
(\delta v)^2\sim \left({gh_{\rm hot}\over \Delta}\right)^2{1\over
f^2+(1/t_{\rm fr}+\eta/t_n)^2}.
\label{pyth}
\end{equation}
Therefore, the component of the shear flow across the front is just 
\begin{equation}
\delta\vec{v}\cdot \hat{n}\sim
\frac{gh_{\rm hot}(1/t_{\rm fr}+\eta/t_n)}{ [f^2+(1/t_{\rm fr}+\eta/t_n)^2] \Delta }.
\label{deltavsh}
\end{equation}
As argued in \S~\ref{sec:horiz-balance-rossby}, the speed of the
front is determined by the speed of the shear flow across the front line,
i.e., $v_{\rm flame}\sim \delta\vec{v}\cdot \hat{n}$. Then, using
$\Delta\sim v_{\rm flame}t_n$ in Eq.~(\ref{deltavsh}), we get
\begin{equation}
v_{\rm flame}\sim \left(
\frac{gh_{\rm hot}}{ t_n}
\frac{1/t_{\rm fr}+\eta/t_n}
{f^2+(1/t_{\rm fr}+\eta/t_n)^2}
\right)^{1/2},
\label{vflame21}
\end{equation}
 
The  typical parameter values relevant for the
burning ocean are $g=2\times
10^{14}\hbox{~cm~s}^{-2}$, $h_{\rm hot}=10^3$~cm, $t_n=0.1$~s, and
$f=2000\hbox{~rad~s}^{-1}$. For purposes of illustration, we take  the eddy
velocity to be equal to the typical convective velocity infered from $1$-D models
of x-ray bursts $v_{\rm turb}=v_c=5\times 10^6\hbox{~cm~s}^{-1}$, and set 
the frictional timescale $t_{\rm fr}$ to $h_{\rm hot}/v_{\rm turb}$. 
We then  obtain the numerical value of 
$v_{\rm {flame}}\sim 200\hbox{~km~s}^{-1}$.
This corresponds to a burst rise time (equator to pole spreading) of
$\sim 0.1$~s.

\begin{figure}
\centerline{\psfig{file=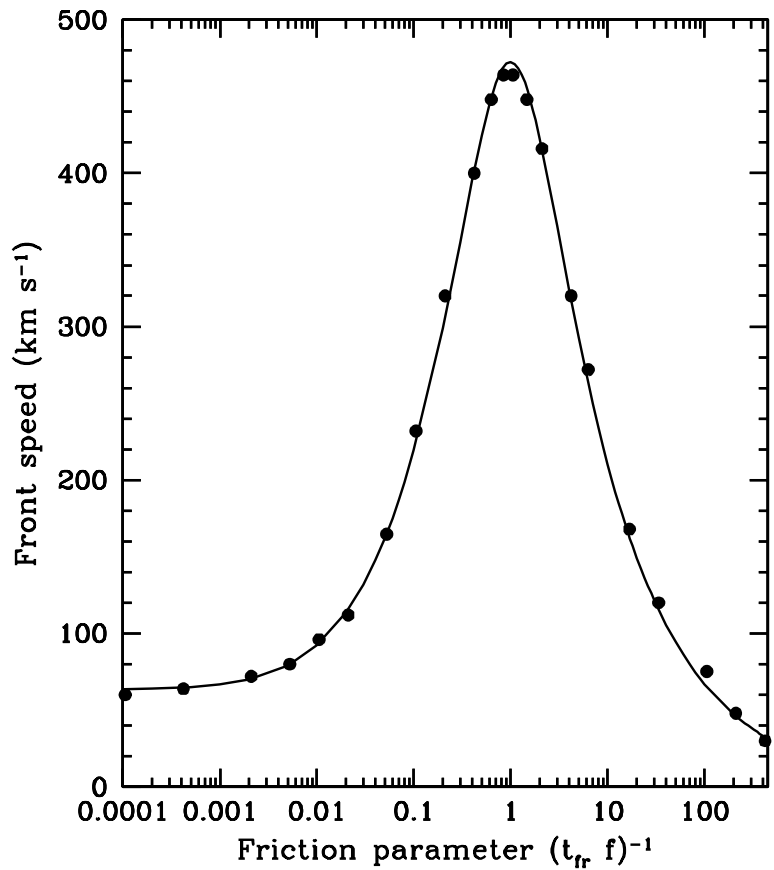,width=4.in}}
%\epsscale{0.9}\plotone{friction.eps}\epsscale{1.0}
\caption{ Front speed as a function of friction strength, $(t_{\rm fr} f)^{-1}$. Dots represent the results of the shallow-water simulation 
(\S \ref{sec:friction}) and the solid line is the fit using the analytical 
expression for the front speed (equation (\ref{vflame21}), \S \ref{sec:horiz-balance-convective}).
}
\label{speedfrict}
\end{figure}

In Figure~\ref{speedfrict} we show the speed of the flame front
$v_{\rm flame}$  as a function of the strength
of the dimensionless frictional coupling constant, $(t_{fr} f)^{-1}$.
The dots represent the results of our model simulations, which 
are discussed in the next section. The continuous line
is a fit made using the functional form of Eq.~(\ref{vflame21}) and
adjusting the values of $h_{\rm hot}$ and $t_n$ to match the
simulation. The agreement between the model simulations and the
analytical formula for the speed propagation is 
very good, which means that the analytical estimates capture the
physics contained in the problem quite well. 

In Figure~\ref{speedfrict}, three regimes of the front speed are
evident. When frictional coupling is very weak (i.e., $t_{\rm fr}\gg
t_n$ and $t_{\rm fr}\gg 1/f$) the front speed asymoptotes to a
constant value given by Eq.~(\ref{vfront12}), i.e., 
we recover the result of
\S~\ref{sec:horiz-balance-rossby} for frictionless flame propagation.
When friction is important (i.e., $t_{\rm fr}\lesssim t_n$), there are
two distinct regimes, 
depending on whether the coupling frequency $1/t_{\rm fr}$ is
smaller or greater than the Coriolis parameter $f$.  When $1/t_{\rm fr}\ll f$
(friction is small), then
\begin{equation}
v_{\rm flame} \sim \left(
\frac{gh_{\rm hot}}{f t_n }\frac{1/t_{\rm fr}+\eta/t_n}{ f}
\right)^{1/2},
\label{vflame22}
\end{equation}
while if friction is strong ($1/t_{\rm fr}\gg f$)
\begin{equation}
v_{\rm flame}\sim \left(
\frac{gh_{\rm hot}t_{\rm fr}}{t_n}\right)^{1/2}.
\label{vflame23}
\end{equation}
For a fixed frictional coupling timescale $t_{\rm fr}$, the front speed is always a
decreasing function of $f$; for a fixed $f$ the front speed attains a
maximum value when friction is acting on the rotation timescale, 
$1/t_{\rm fr}= f$:
\begin{equation}
v_{\rm flame}^{\rm max}\sim \left(gh_{\rm hot}\over ft_n\right)^{1/2}.
\label{vflame24}
\end{equation}

Comparison with Eq.~(\ref{vfront12}) shows that the maximum speed is
a factor of $(ft_n)^{1/2}\sim 15$ larger than the front speed without
friction. The increase of the front speed when
 friction is introduced into the ocean can 
be qualitatively understood as follows.
Without
friction, the dominant component of the fluid velocity, $v_{\rm
geostrophic}$, is exactly parallel to the front, and the cross-front
velocity, $\delta v_{\rm ageostrophic}$, is a factor of $f t_n$
smaller.  The presence of friction modifies geostrophic balance so
that the geostrophic shear flow  has a component perpendicular to the
front.  However,
if  friction becomes too large, it will suppress the
shear, and the front will stall. The cross-front component of 
geostrophic shear is maximized when friction and
Coriolis force have  the same magnitude, $1/t_{\rm fr}\approx f$.

In the above analytical estimates we neglected the effect of the
magnetic field that may be present on the neutron star surface. What
is the magnitude of a $B$ field that would have dynamical consquences?
This $B$ field would have to alter the leading-order geostrophic
balance [cf. Eqs~(\ref{forcebalance1}) and~(\ref{forcebalance11})],
i.e., $B_{\rm dynamical}\sim (4\pi\rho v_{\rm geostrophic}^2)^{1/2}$.
If viscous coupling is negligible, $v_{\rm geostrophic}\sim(gh_{\rm
hot})^{1/2}$, and 
\begin{equation}
B_{\rm dynamical}\sim 10^{12}{\rm~G~}
\left(\frac{\rho}{10^6{\rm~g~cm}^{-3}}\right)^{1/2}
\left(\frac{h_{\rm hot}}{10^3{\rm~cm}}\right)^{1/2},
\end{equation} 
i.e., magnetic energy density needs to be comparable to the
hydrostatic pressure in order to affect geostrophic flow. Conversely,
if convective viscosity is such that $v_{\rm flame}=v_{\rm flame}^{\rm
max}$, then $v_{\rm geostrophic}\sim gh/f\Delta\sim (g h_{\rm
hot}/ft_n)^{1/2}$, and the above estimate is a factor of
$(ft_n)^{1/2}\sim15$ smaller.  Even in order to affect the first-order
correction to the geostrophic balance, i.e., the cross-front
circulation, the magnetic pressure has to be comparable to $\rho
v_{\rm ageostrophic}^2/2$. The smallest ageostrophic velocity is
attained in the case of no friction, $v_{\rm ageostrophic}\sim (gh_{\rm
hot})^{1/2}/ft_n$ [cf. Eqs~(\ref{vfront11}) and (\ref{vfront12})],
which still requires $B_{\rm dynamical}\gtrsim 10^9$~G in order to
affect the details of the cross-front circulation.  As discussed in
the introduction, it is unlikely that accreting neutron stars in LMXBs
possess surface magnetic fields of this magnitude. 

The argument above applies to the instanteneous effect of magnetic
field on the flow, and does not take into account the nonlinear
evolution of magnetic field in the (possibly turbulent) shear flow,
and the back-reaction of the evolved field on the flow (see, e.g.,
Kluzniak and Ruderman 1998, Spruit 1999, Cumming and Bildsten
2000). Detailed consideration of this issue is beyond the scope of
this paper. In Section $4$ we only briefly speculate on magnetic field
generation by an MHD dynamo in the burning front.

In the next section we construct an explicit $2$-layer shallow-water
model for flame propagation, which we designed to include the
essential physics described in this section. Our numerical
simulations confirm the analytical results presented here.

\section{Propagation of the flame: numerical model}\label{numerical}
\subsection{Motivation and basic description of the two-layer model}

In this section we outline one particular approach to modeling hydrodynamic
flows in a stratified plane-parallel atmosphere\footnote{Henceforth we use the
terms ``atmosphere'' and ``ocean'' interchangeably.}  in the presence of
localized heat sources and sinks (thermal forcing). The thermodynamic state of a
parcel of fluid in the atmosphere will be described using its temperature, $T$,
pressure, $p$, and potential temperature, $\theta=T (p_0/p)^\kappa$, where
$p_0$ is some constant reference pressure, and $\kappa$
is related to the
adiabatic index of the gas, $\gamma$, as $\kappa=1-1/\gamma$. Potential
temperature can be interpreted 
 as  the temperature which a parcel would have if it were
adiabatically brought from pressure $p$ to the reference pressure $p_0$
(Holton, 1992). It equals to $\exp (s)$, where
$s$ is the specific entropy of the parcel.
In a stably stratified atmosphere potential temperature should
increase or be constant with height. Another property which follows from the
definition of $\theta$ is that for fixed pressure 
it is proportional to the specific volume of the
gas: $\theta \propto {1\over \rho} p^{(1-\kappa)}$. For the particular case
$\kappa=1$ which will be utilized below, potential temperature is equivalent to
the specific volume of a parcel of fluid.

\begin{figure}
\centerline{ \psfig{file=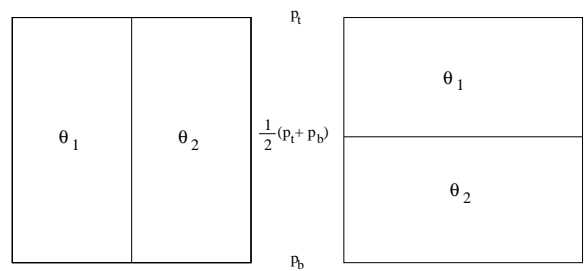,width=3in}}
\vskip .1in
\caption{
Adjustment to equilibrium in a fluid with an entropy gradient. a) Initial 
configuration with two columns of equal mass with uniform potential temperature
$\theta_1>\theta_2$ in vertical hydrostatic balance.
Upper surfaces are at the same external pressure $p_t$, but not at the
same height;
b) Final configuration -- entropy stratification. 
Higher entropy fluid is on top, and the interface is at pressure ${1\over 2} (p_t+p_b)$.}
\label{fig:columns}
\end{figure}

Now we can begin to evaluate the 
qualitative effects of heating on an atmospheric
column. From the heat equation we have:
\begin{equation}
{d \ln \theta \over dt}= {J \over c_p T},
\label{heateq}
\end{equation}
where $J$ is the net rate of heating (or cooling) per unit mass,
which, in our case, is the difference between the nuclear energy
generation rate and the radiative cooling rate. The effect of heat
input, therefore, is to change the potential temperature of the
element of fluid. Such a change, coupled with maintenance of vertical
hydrostatic balance in the column leads to hydrodynamic circulation.
As a simple example of how such a circulation arises consider two
columns of gas of equal column density but of different potential
temperature ($\theta_1>\theta_2$) positioned next to each other (see
Fig. \ref{fig:columns}a).  Each of the columns is assumed to be in
vertical hydrostatic equilibrium. This situation can arise if column 1
is strongly heated, so that its entropy is instantly increased
relative to the entropy of column 2.  This configuration, however, is
not stable: the fluid with higher entropy will end up on top of the
fluid with lower entropy, and the atmosphere will become stably
stratified (see Fig. \ref{fig:columns}b).  The dynamics of this
adjustment can be qualitatively described as follows.  Initially, let
the tops of the columns be at the same outside pressure; the bottom
pressure then will also be the same since the column densities are equal
for both columns.  At higher altitude the pressure in column $1$ is
larger than in column $2$ at the same height because the density in
column $1$ is lower than that of column $2$ (cf. Fig.~\ref{figfront}).
At first the fluid will flow
from column $1$ towards column $2$ along the top, and not along
the bottom where the 
pressures are the same.  As the high-entropy fluid accumulates
on top of column $2$
the pressures
at the bottom no longer match, and there will be a flow of fluid from column
$2$ towards column $1$ along the bottom as well. This circulation will continue
until the equilibrium state shown on Fig. \ref{fig:columns}b is achieved.
Note that we assumed this equilibration is happening adiabatically, so that
the fluids of different entropy maintain their identity and in the end do not
mix but stratify.

The entropy stratification described above suggests a useful analogy for the
effect of a heat source in the atmosphere. Since higher entropy fluid ends up
on top of the lower entropy fluid, we can treat entropy (or potential
temperature) as a generalized vertical coordinate. The action of the
heat source is then
to pump fluid from a low entropy level to a high entropy level
while conserving mass. The isentropic levels generally do not coincide with the
levels of constant height, and there will be a circulation of fluid.  If the
motion outside the heat source is adiabatic, it is impossible
for the fluid to cross the isentropic level again, and the motion will be along
the isentropes.  

Henceforth, we will adopt a particularly simple, yet powerful model 
which uses the equation of state with $\kappa=1$ ($\gamma\to
\infty$). 
The fluid with such an equation of state is quasi-incompressible:
it does not allow adiabatic motions which involve compression and rarefaction, 
and has an infinite speed of sound. However, for $\kappa=1$, the 
potential temperature is equal to the specific volume of the fluid,
and heating will lower the density of the fluid while conserving its mass.

 Following Ooyama (1969) we consider an atmosphere consisting of two layers of
incompressible fluid of different density, with density ratio $\epsilon \equiv
\rho_2/\rho_1 < 1$.  The lighter fluid is assumed to be in the top layer (layer
2).  A fluid element is allowed to be in one of the two possible density
states, $\rho_1$ or $\rho_2$. The heating is assumed to be located at the
interface between the two layers (which is not necessarily a surface of
constant height); the physical effect of the heating is to convert the ``cold''
dense material of density $\rho_1$ into the ``hot'' light material of density
$\rho_2$, transferring the fluid from the lower (cold) to the upper (hot)
layer.  
The equations of motion for the two layers are then given by the shallow-water
system of equations (Pedlosky 1987) individually formulated for each layer.
The crucial distinction of our model is the interlayer 
coupling source/sink term in the mass
continuity equation, and the friction between the layers:
\begin{eqnarray}
{\partial h_1\over \partial t} + {\mathbf \nabla}_{||}\cdot h_1 {\mathbf v}_1 &=& -\epsilon({Q}_{+}-{Q}_{-}), \label{sys1} \\
{\partial h_2\over \partial t} + {\mathbf \nabla}_{||}\cdot h_2 {\mathbf v}_2 &=& {{Q}_{+}-{Q}_{-}}, 
\label{cont2}\\
{\partial v_{1x}\over \partial t} +  {\mathbf v}_1 \cdot {\mathbf \nabla}_{||}  v_{1x} &=& - g{\mathbf \nabla}_{x} (h_1+
\epsilon h_2)+{f} v_{1y} \nonumber \\
&-&{\epsilon\over h_1}\left({Q}_{-}+\mu_f\right)(v_{1x}-v_{2x}), \label{mom1}\\
{\partial v_{2x}\over \partial t} +  {\mathbf v}_2 \cdot {\mathbf \nabla}_{||}  v_{2x} &=& - g{\mathbf \nabla}_{x} (h_1+
 h_2)+{f} v_{2y} \nonumber \\
&-&{{Q}_{+}+\mu_f\over h_2}(v_{2x}-v_{1x}),\\
{\partial v_{1y}\over \partial t} +  {\mathbf v}_1 \cdot {\mathbf \nabla}_{||}  v_{1y} &=& -{f} v_{1x}  %\nonumber \\
-
{\epsilon\over h_1}\left({Q}_{-}+\mu_f\right)(v_{1y}-v_{2y}),\\
{\partial v_{2y}\over \partial t} +  {\mathbf v}_2 \cdot {\mathbf \nabla}_{||}  v_{2y} &=& -{f} v_{2x} %\nonumber \\
-
{{Q}_{+}+\mu_f\over h_2}(v_{2y}-v_{1y})
 \label{sys1end}, 
\end{eqnarray}
cf.~Eqs (3.5)--(3.7) of Ooyama (1969).  Here, $h_{1,2}$ are the
heights of the cold
and hot layers respectively.  We consider the dynamics on an $x,y$ plane with
${\mathbf \nabla}_{||}= ( \partial/\partial x, \partial/\partial y)$.  To
keep the problem tractable we will allow variation only in $x$ direction, so
that we have a slab geometry in $x$ with possible tangential velocity $v_y$.
Depending on the case under consideration, the plane of the simulation will be
oriented in various ways on the neutron star surface.  The heating rate $Q
\equiv J/g$ represents the flux of material from the lower to the upper layer, and has
dimensions of velocity (for the derivation of Eqs. (\ref{sys1}---\ref{cont2})
see the appendix).  
The quantities ${Q}_{+}$ and ${Q}_{-}$ are defined as follows:
${Q_+}={Q}$, ${Q_-}=0$ if the heating rate ${Q}>0$, and ${Q_-}=-{Q}$, ${Q_+}=0$
if ${Q}<0$. The terms on the right-hand side of the above equations which are
proportional to $\mu_f(v_1-v_2)$ and $Q_{\pm}(v_1-v_2)$ represent the friction
between the two layers, the former due to (perhaps convective) viscosity, and
the latter due to momentum-conserving transport of material from one layer to
another.

In general, the heating rate depends on the temperature at the point
where the heat source is applied.  Ooyama (1969) argues that for the
shallow water model it is reasonable to assume that the temperature
of the fluid can be represented by its height. His argument is roughly
as follows: an adiabatic temperature rise in a normal fluid happens
when the flow is converging; but a converging flow in a shallow-water
fluid results in a rise of its height. Likewise, an increase
of the temperature due to heat input corresponds to addition of fluid
to a shallow-water layer, and hence corresponds to an increase of its
height. Thus the  temperature increase in a
normal fluid, both due to compression
and due to the heat input, corresponds to the 
height increase in a shallow-water fluid.
In our model, we shall assume that the temperature at the interface
where the heating occurs is set by the height of the second, light
layer through a simple relationship $T=g h_2/c_p$.

In the following three subsections we consider how our model responds to
localized sources of heat.  
This will help us develop intuition which will be useful in
analyzing the ignition and spreading of the burning front.

\subsection{Localized heating in a nonrotating atmosphere without friction}

In a nonrotating atmosphere the effective Coriolis parameter $f$ is
zero, and this may be interpreted as the burning in a region along the
rotational equator of a neutron star. Consider a part of the atmosphere of
a large lateral extent with the two-layer fluid described above in the
equilibrium configuration with, for example, $h_1=h_2$ initially.  The fluid is
allowed to freely leave and enter the domain through the sides. For a
delta-function heat source $Q=Q_0 \delta(x)$ and zero friction between the
layers ($\mu_f=0$) we can find a steady state solution that will be achieved
after an initial transient.  The solution will have a uniform divergent flow at
the upper layer, satisfying $h_2 v_{2x}= {Q_0} (\Theta(x)-1/2)$, and a uniform
convergent flow at the bottom layer, with $h_1 v_{1x}|=- \epsilon Q_0
(\Theta(x)-1/2)$, where $\Theta(x)$ is a step-function. The equilibrium value
of $h_2$, after the transient flow traverses the domain of interest, is larger
by $\sim Q_0/\sqrt{gh}$ than the initial thickness of the upper layer, while
the layer 1 is contracted by similar amount.  The main property of this
solution is that, even though the heat source is continually operating, the
temperature (or thickness) of the layers does not change in time after the
initial transient has died out. The heating  drives opposing winds in 
 both layers, and the heating  energy goes into  raising the fluid from the 
bottom to the top. For small heating rates, $Q_0/h \lesssim \sqrt{gh}$, the shearing
flow between the two layers is stable to Kelvin-Helmholtz instability due to
buoyant stratification.
   
This solution is valid until the outflow has reached the end of the lateral
extent of the atmosphere, at which point layer 2 will grow at the expense of
depleting layer 1. Such lowering of layer 2, i.e., the presence of high-entropy
fluid at a lower height, represents simple heating of the whole atmosphere that
we eventually expect from a heat source.  Therefore, in order to have the heat
source continuously increase the temperature locally, as is required for a local
thermonuclear runaway, one needs a way of containing the outflow over some
spatial length scale. In the absence of lateral boundaries, this feat is
accomplished by rotation of the star, friction, or both.

\subsection{Localized heating in a rotating atmosphere }\label{locheat}
\paragraph{General considerations.}
The main new feature introduced into the dynamics in a rotating atmosphere is
that when friction is absent, all disturbances do not relax to a state of
minimum potential energy, in contrast to a nonrotating atmosphere.  Instead,
 disturbances tend to relax rapidly (in a time of the order of one rotation
period) to a state of geostrophic balance, where all pressure gradients are
balanced by the Coriolis force acting on fluid moving along the isobars.  A
famous example is known as the Rossby adjustment problem (Gill~1982): a
uniformly rotating fluid of constant density is released with an initial step
in its height distribution.  A transient ensues, in which the fluid starts to
spread. The spread is slowed down, however, by the action of the Coriolis
force, and the fluid oscillates around the equilibrium configuration,
radiating gravity waves.  The end result is a state of equilibrium with nonzero
kinetic energy in the flow such that the adjusted pressure gradient is
balanced by the Coriolis force. A simple way of understanding the spatial
scale on which the equilibration occurs is to consider the momentum equations
for a single incompressible layer of constant depth $H_0$ with a perturbation
$h=H_0+\delta h$:
\begin{eqnarray}
{d v_x \over d t}&=&  -g {\partial \over \partial x} (H_0+\delta h) + f v_y, \label{long} \\
{d v_y \over d t}&=& -f v_x.\label{trans}
\end{eqnarray}
Equation~(\ref{trans}) can be integrated to give $v_y=f \Delta x$, i.e.,
the transverse velocity of the fluid element is proportional to the
displacement from its initial position.  Substituting this into (\ref{long})
with the assumption of geostrophic balance, and estimating the pressure
gradient as $g H_0/\Delta x$, we obtain the characteristic length over which
the pressure gradient is spread: $\Delta x = \sqrt{g H_0}/f$. This is the
Rossby radius of deformation for a rotating atmosphere of scaleheight
$H_0$. After the transient, the initial step in height of the fluid will
transform into a gradual slope over the lengthscale of Rossby radius.

The above argument is valid only if the horizontal displacement
of the fluid element, $\Delta x$, is of the same order of magnitude
as the lengthscale over which the pressure gradient is spread. This
is true only for large initial perturbation, $\delta h\sim H_0$.
However, the conclusion that the
pressure gradient is spread over the Rossby radius turns
out to be valid even if the initial perturbation is not
large. For a small perturbation, the spread of the
pressure gradient does not involve actual motion of a fluid element over 
$\Delta x$, but is communicated via gravity waves.
In the presence of 
rotation, the gravity waves are dispersive, with dispersion relation
$\omega^2=f^2+c_g^2 k^2$, where $c_g=\sqrt{g H_0}$. For wavelengths shorter
than the Rossby radius, gravity waves are unaffected by rotation and leave the 
region of disturbance, while for wavelengths  
longer than $c_g/f$ the propagation speeds become very small. 
Only the long-wavelength component of the initial perturbation is 
left behind, and even for small initial perturbations the geostrophic 
ballance is eventually established 
on the lengthscale of the Rossby radius, $\sqrt{g H_0}/f$. We refer the reader
to Gill (1982) for further details. 

\paragraph{Localized heating in a rotating atmosphere without friction.}
We now study the  flows in
 a rotating atmosphere forced by a localized heat input. We
 concentrate on the top layer of the isentropic model
 discussed above.  If we assume that the density contrast between the two
 layers is large enough ($\epsilon \ll 1$) then the height of level 1 does not
 change appreciably during the evolution. In that case, the linearized
 equations for the top layer of thickness $h=H_0+\delta h$ become:
\begin{eqnarray}
{\partial h\over \partial t} &+& H_0 {\partial \over \partial x}v_x = Q(x,t), \label{cont1} \\
{\partial v_{x}\over \partial t} &=& -g{\partial \over \partial x} h+{f} v_{y}, \label{xmoment}\\
{\partial v_{y}\over \partial t} &=& -f v_x. \label{ymoment}
\end{eqnarray}
We will further refer
to models tracking only the top fluid layer as one-layer models.  We find that
they contain most of the essential physics of the two-layer system.  If the
timescale of heating is much slower than the time it takes gravity waves to
propagate over the characteristic Rossby radius (i.e., $1/f$), at each moment in
time the system will be in approximate geostrophic balance.  We thus omit the
time derivative in (\ref{xmoment}); this also automatically filters out
transient gravity waves.  The system of equations can now be combined to yield
a single equation for the height of the layer:
\begin{equation}
h-a_R^2 {\partial^2 h\over \partial x^2}  = \int_0^t Q(x,t') dt'. \label{kleingord}
\end{equation}
This is a forced Klein-Gordon equation, where the source term is provided by
the total heating at a particular point.  The Green's function for the left
hand side of the equation is $G(x|\zeta)={1/2 a_R} e^{-|x-\zeta|/a_R}$. Therefore,
for a constant delta-function heating $Q(x,t)=Q_0 \delta(x)$ the fluid  follows a
growing vortex sheet solution:
\begin{eqnarray}
h(x,t)&=&{Q_0 t \over 2 a_R} e^{-|x|/a_R},\\
v_x(x,t)&=&{\rm{sign}}(x) {Q_0 \over 2 H_0} e^{-|x|/a_R},\\
v_y(x,t)&=&-{\rm{sign}}(x) {Q_0 t f\over 2 H_0}e^{-|x|/a_R}.
\end{eqnarray}
A few points should be made about this solution. It shows that, for constant
heating, the height of the layer (or its temperature) grows linearly in time in
a region of characteristic size equal to the Rossby radius around the heat
source. Aside from nonlinear effects, the size of the area affected by heating
does not change as the height increases. Therefore, the pressure gradient
increases linearly with time.  The velocity transverse to the pressure gradient
(geostrophic velocity $v_y$) also grows linearly with time. If we were to solve
the same problem with cylindrically symmetric localized heating, the
geostrophic flow would create a vortex-like circulation around the source of
heating.  When $Q_0>0$ the heated region is the area of high pressure, and
hence the sense of geostrophic velocity in the layer is anticyclonic (opposite
to the sense of rotation of the star), and cyclonic in a low pressure region
for the case of cooling.  The discontinuity in velocity at $x=0$ is an
artifact of the delta-function forcing.  Since realistic forcing is 
distributed over a finite area, the velocity should  smoothly go to zero at the
center of heating.  Another feature of the solution is the presence of
a nonzero
flow away from the source of heating (and the opposite for cooling). This flow
is not due to pressure gradient as such, but rather due to a time rate of
change in the pressure gradient as the central pressure rises or falls. Such a
flow is known as the {\it isallobaric wind} (Holton, 1992) and is not itself in
geostrophic balance. 
However, this wind plays an important role in the adjustment to 
balance as can be
seen from Eq.~(\ref{ymoment}). In order for the geostrophic velocity to change
in response to modified conditions, there must be a nonzero ageostrophic
velocity; cf.~Eq.~(\ref{thermalwind2}).  Since all quantities in the above
solution exponentially decay at distances larger than the Rossby radius, the
effects of heating are thus localized due to rotation.

\paragraph{Localized heating in a rotating atmosphere with strong friction.}
When there is (turbulent) viscosity in the system, frictional
forces need to be added to the right-hand side of Eqs. (\ref{xmoment}) and
(\ref{ymoment}):

\begin{eqnarray}
{\partial h\over \partial t} &+& H_0 {\partial \over \partial x}v_x = Q(x,t), \label{cont11} \\
{\partial v_{x}\over \partial t} &=& -g{\partial \over \partial x} h+{f} v_{y}-{v_x\over
 t_{\rm fr}}, \label{xmoment1}\\
{\partial v_{y}\over \partial t} &=& -f v_x-{v_y\over t_{\rm fr}}. \label{ymoment1}
\end{eqnarray}
Here $1/t_{\rm fr}$ 
is the coefficient of frictional drag (cf. \S~\ref{sec:horiz-balance-convective}). 
For $1/t_{\rm fr}, f\gg
1/t_n$ we can neglect the time derivatives on the left-hand side of Eqs.~(\ref{xmoment1}) 
and (\ref{ymoment1}).  We then get
\begin{eqnarray}
{\partial h\over \partial t}&+&D{\partial^2h\over\partial x^2}=Q(x,t),\label{diffusion}\\
v_x&=&-{D\over H_0}{\partial h\over \partial x}, \label{diff1}\\
v_y&=&-{ft_{\rm fr}}v_x, \label{diff2}
\end{eqnarray}
where 
\begin{equation}
D={t_{\rm fr}gH_0\over 1+(ft_{\rm fr})^2}.
\label{diffcoeff}
\end{equation}
Equation (\ref{diffusion}) is a diffusion equation with the source term $Q$ and
the diffusion coefficient $D$. For a localized heating $Q(x, t)=Q_0\delta (x)$
switched on at $t=0$, the solution is given by
\begin{equation}
h(x, t)=Q_0\sqrt{t\over 4\pi D}F(x/2\sqrt{Dt}),
\label{diffsolution}
\end{equation}
where 
\begin{equation}
F(q)=q\int_q^{\infty}{1\over q_1^2}e^{-q_1^2}dq_1.
\end{equation}
The height (i.e.~temperature) and velocity perturbations are concentrated within
 the diffusion length $\delta x=\sqrt{Dt}$.

\paragraph{Implications for burst ignition lengthscale.}
What can we learn from the atmospheric response to localized heating?  We saw
that for a frictionless rotating atmosphere, the temperature perturbation is
confined to the Rossby adjustment radius, hence the ignition of the burst is
likely to happen on that scale. When there is strong friction, the temperature
perturbation is confined to the diffusion lengthscale. In this case, we
therefore expect the ignition to happen on the scale $\delta x_{\rm
ignition}\sim \sqrt{Dt_n}$, where $D$ is the diffusion coefficient given by
Eq.~(\ref{diffcoeff}), and $t_n$ is the characteristic nuclear burning
timescale. As we saw in the previous section, and as will be confirmed by
simulations in the next section, these ignition lengthscales are of the same
order as front widths in respective cases; cf.~Eqs (\ref{width2}) and
(\ref{vflame21}). The only difference is that in the case of the front
propagation the important scaleheight is that of the hot part of the ocean,
whereas for the ignition the scaleheight $H_0$ is that of the unheated, cold
ocean.

\subsection{Burning front propagation}

\begin{figure}
\centerline{\psfig{file=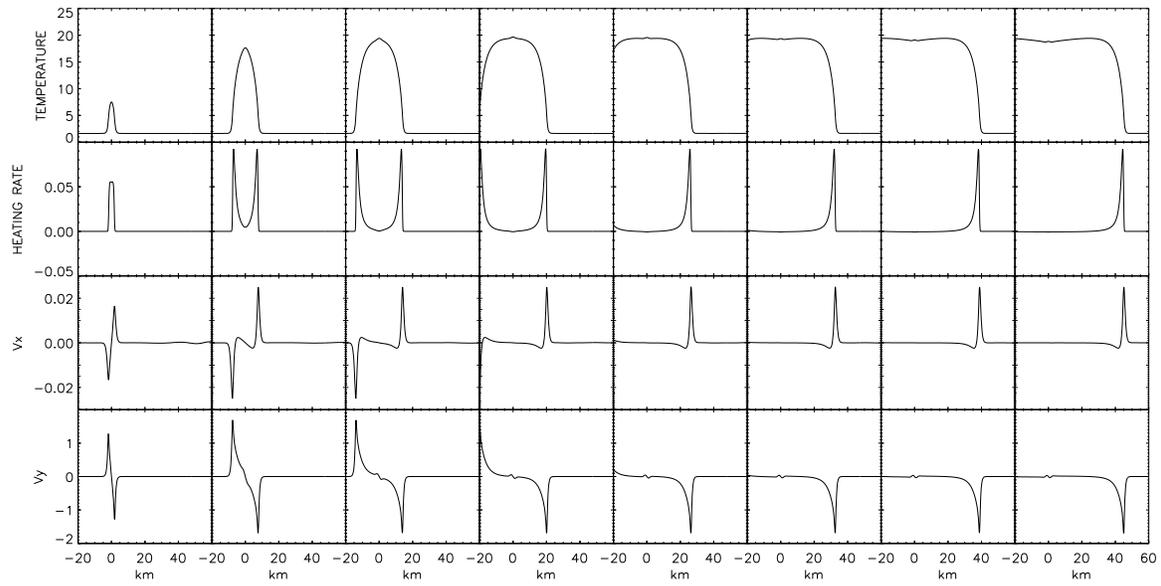,width=6in}}
\caption{ One-layer model of the evolution of a localized source of heating into two propagating
 burning fronts.  Shown (from top to bottom): temperature (scaleheight) of the layer in units
of $10^8$ K; 
instantaneous heating rate (in units of $5\times 10^{19} ~{\rm erg ~g^{-1} s^{-1}}$); 
cross-front ageostrophic velocity $v_x$  and geostrophic velocity $v_y$ 
parallel to the front (velocities are in units of gravity wave speed
of the cold material, $1.5 \times 10^8 {\rm ~cm ~s^{-1}}$).
Time is increasing left to right with frames separated by $0.125$ sec.
}
\label{fig:spread}
\end{figure}

When we numerically solve the system of equations (\ref{sys1})-(\ref{sys1end})
with a temperature-sensitive heating function, we find that a local runaway can
develop into a propagating burning front solution. In order to simulate
conditions relevant to the case of a neutron star atmosphere during a burst we
consider heating due to $3\alpha$ helium-burning reaction and one-zone cooling
due to black-body radiation (see, e.g., Cumming and Bildsten 2000)\footnote{
The heating function that we use is representative of the conditions in a type
I burst, but is by no means complete. In particular, 
we neglected the electron screening of the $3\alpha$ reaction 
(Fushiki and Lamb, 1987), as well as further energy release due to nuclear 
evolution beyond carbon. We find that the physics of front propagation is 
independent of the
particular heating function as long as this function has a rise, a single
peak, and decays at high temperatures.  
}:
\begin{equation}
Q=5.3\times 10^{21} {\rm erg ~g^{-1} ~s^{-1}} {\rho_5^2 Y^3 \over T_8^3} \exp \Big({-44\over T_8}\Big)-
{a c T^4 \over 3 \kappa y^2}. \label{heatfunc}
\end{equation}
Here, $\rho_5$ is the density in units of $10^5 \hbox{~g~cm}^{-3}$ (which we
evaluate including degenerate corrections), $Y$ is the helium abundance obeying
$dY/dt=\epsilon_{3\alpha}/(5.84\times 10^{17} {\hbox{ ~erg/s}})$, $T_8$ is the
temperature in units of $10^8$ K, and $\kappa$ is the opacity, which, for
simplicity, we take to be a constant, $0.03{\rm ~cm^{2}~g^{-1}}$.
 For our initial
state we consider the atmosphere of pure helium 
with column depth $y=5.4\times 10^8$~g~cm$^{-2}$ at 
the point of ignition. The temperature at the start of the runaway
(determined as the point where temperature derivatives of heating and cooling
match) is $T_8=1.64$. The height of the top fluid layer, which in our model
represents temperature, is normalized in units of the scaleheight of the
unperturbed atmosphere. In order to set off the runaway we raise the
temperature at one location on our grid on a scale much smaller than the Rossby
radius.  The sequence of snapshots of subsequent evolution
of layer height (temperature), instantaneous net heating $Q(x,t)$, and
ageostrophic and geostrophic velocities in the top layer is shown in
Fig. \ref{fig:spread} in the one-layer limit of our model ($\epsilon= 0$).\footnote{ 
For our numerical method we use second order accurate
finite-differencing of model equations.  This allows us 
to cut down on numerical
diffusion that can artificially increase front speeds. Since the underlying
equations support fast moving dispersive gravity waves, special attention is
paid to boundary conditions. To allow for integration times longer than the
gravity wave crossing time, we introduce an absorbing boundary layer (Romate
1992) that dissipates the gravity waves of wavelength smaller than the size of the
boundary layer. To prevent reflection of longer wavelengths we add the
non-reflecting condition that eliminates backward going characteristics at the
boundary (Thompson 1987).}
 In
this simulation, 
the Coriolis
parameter is constant everywhere on the grid and is representative of a star
with $250$ Hz rotation frequency at $45^\circ$ latitude.  First we discuss
simulations for the case without interlayer friction [$\mu_f=0$ in
(\ref{sys1}-\ref{sys1end})].

 The initial height perturbation 
introduces
potential energy about 2/3 of which is radiated into gravity waves, and the
rest ends up in a geostrophically balanced flow which is established within a
rotation period and has a characteristic lengthscale of a Rossby radius.  As
expected, the initial stage of growth is similar to the delta-function response
described above: the vortex intensifies while maintaining its width.  Since in
the beginning of the runaway the heating is a strongly increasing function of
temperature, the dominant contribution to vortex intensification comes from the
fastest-growing central part.  
The
heating function (\ref{heatfunc}) goes through a peak during the runaway at
about $9\times 10^8$ K for our parameters before cooling catches up with
heating at about $2\times 10^9$ K. When the central temperature moves past the
peak heating, and the tails of the vortex start undergoing the fast part of the
runaway, a double-peaked structure of the heating function appears as seen in
fig. \ref{fig:spread}.  As the heating peaks separate, the vortex begins to
spread and the structure of a burning front appears: a localized area with the
runaway heating which moves into the unburned material at a constant speed. The
burning is accompanied by a rise in layer thickness, so that
there is a substantial pressure gradient in the direction of motion of the
front. In the absence of friction, this pressure
gradient is closely balanced by the Coriolis force acting on the flow generated
parallel to the front. As discussed in Section \ref{secanalyt}, the unbalanced
ageostrophic component is present and necessary to drive the change in the much
larger geostrophic flow: $d v_{\rm {geostrophic}}/dt=-f
v_{\rm {ageostrophic}}$.  It is the ageostrophic flow (typically smaller in
magnitude by a factor of order $1/f t_n\sim 2\times 10^2$ compared to its geostrophic
counterpart) which is responsible for the front propagation.

\subsection{Structure and speed of the front \label{structure}}

The temperature at a location ahead of the front increases due to two effects:
influx of the ageostrophic wind of hot material across the front, and local
thermonuclear energy generation.  The local heating rate due to thermonuclear
reactions, which sensitively depends on the temperature, will start to grow if the
temperature perturbation is sufficient to push the fluid parcel into the
runaway regime.  We are assuming that the temperature and column depth in the
fluid are such that by raising the temperature sufficiently the runaway will
start. 
In this case, since the
temperature well inside the burning front is high enough for a runaway, we are
guaranteed that even if the fluid before the front is colder than the point of
marginal stability, it will cross into the runaway as the front approaches.
When the friction is absent, the fluid motion across the front is driven by the
time rate of change in the pressure gradient: $v_x = - (g/f^2) {d \over d t}
\nabla h$ [cf. eq. (\ref{shear1})].  As the front approaches, the pressure
gradient  increases in magnitude (it is negative for a front moving to the
right) and reaches the maximum at the point of peak heating.  The fluid is thus
pushed through the front, igniting material ahead, and the pressure gradient is
performing work to accelerate this fluid
 to the geostrophic velocity.  The width of
the front is thus set by the magnitude of the Coriolis force which
turns the cross-front fluid motion into the flow parallel to the front.  Since
the scaleheight inside the front can change by factors of order 10 for strong
bursts, nonlinear effects become important.  Typically, however, the
characteristic front thickness is of the order of the Rossby radius in the hot
material behind the front [$\sim 3 \hbox{km}$ for our parameters,
cf. Eq. (\ref{width3}) ]. Since the burning must be complete inside the
front width, the speed of the front should be such that it moves one Rossby
radius in a characteristic nuclear time.  For our parameters, it takes 
$0.1-0.2$
seconds to go through the fast part of the $3\alpha$ burning, which yields an
estimate of $15-30$~km~s$^{-1}$ for the front speed.  Behind the location of
peak heating the temperature continues to rise, but the magnitude of the pressure
gradient starts to decrease, and the isallobaric wind is directed opposite to
the motion of the front.  The relative amount of cross-front fluid motion in
the forward and backward directions can dramatically influence the speed of
propagation of the front and depends on frictional effects due to (turbulent)
viscosity and on drag due to momentum-conserving transport of fluid between
layers. We shall refer to the latter effect as momentum coupling.

\begin{figure}
\centerline{\psfig{file=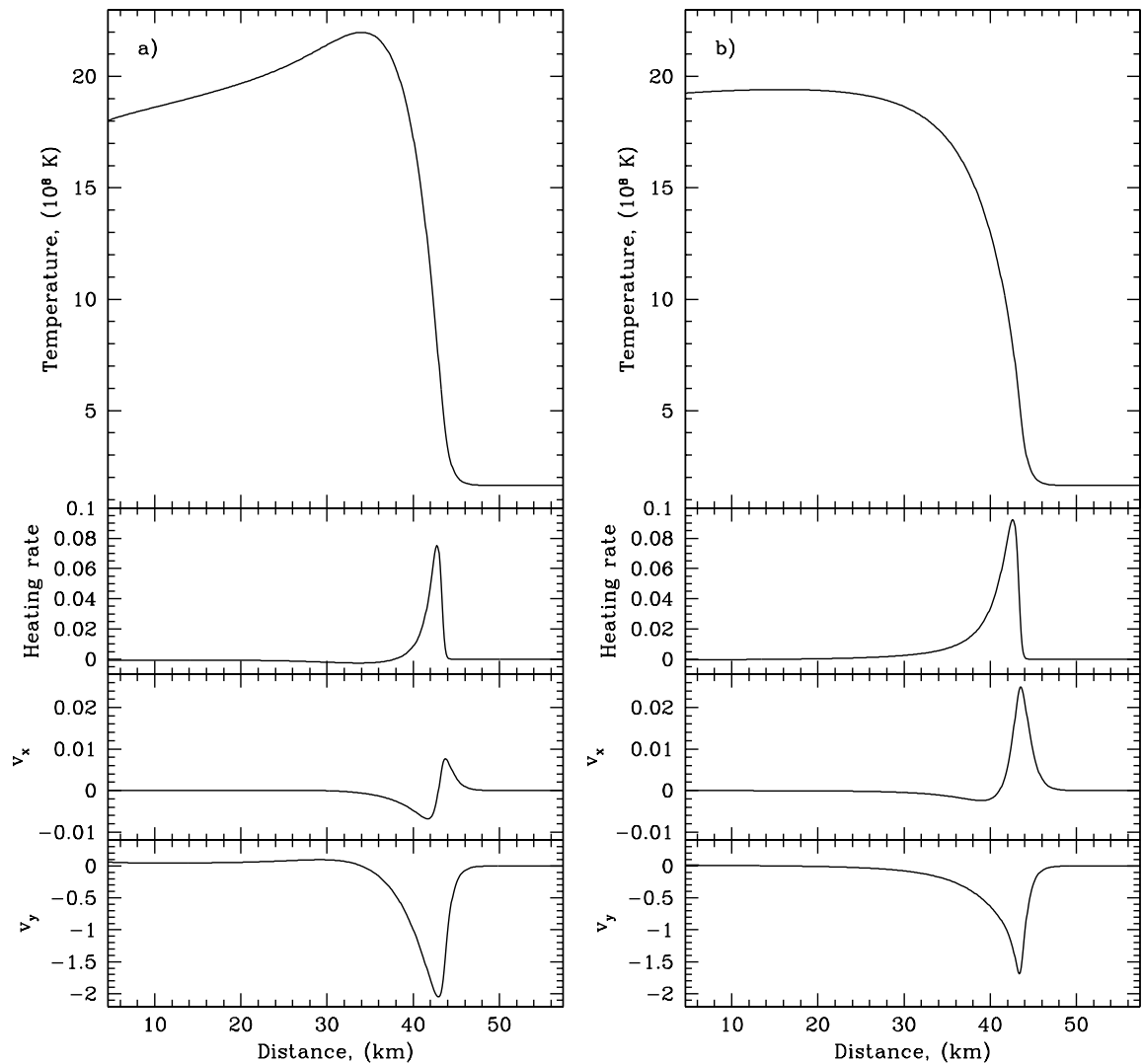,width=6in}}
\caption{ Internal structure of burning fronts in the one-layer model. Shown are the temperature of the layer, 
the instantaneous heating rate
(in units of $5\times 10^{19} ~{\rm erg ~g^{-1} s^{-1}}$), 
cross-front ageostrophic velocity $v_x$, and tangential geostrophic
velocity $v_y$ (velocities in units of gravity wave speed in the cold 
material, $1.5 \times 10^8 {\rm ~cm ~s^{-1}}$).
a) front without the drag due to momentum coupling; b) momentum-coupling drag included.
}
\label{twofronts}
\end{figure}

Figure {\ref{twofronts}} shows the internal structure of two fronts computed in
the one-layer approximation with identical heating functions but differing in the
strength of momentum coupling.  Front A (fig. \ref{twofronts}a) is
computed without the terms proportional to $(v_{2x,y}-v_{1x,y})$ in
Eqs.~(\ref{sys1})--(\ref{sys1end}) (momentum-nonconserving front), while front B
(fig.~\ref{twofronts}b) is computed with momentum-coupling terms
retained.  Viscous friction is turned off in both cases ($\mu_f=0$).
Physically, case A represents the situation where the fluid from the cold layer
is being injected into the hot layer at the velocity of the hot layer, while in
B the fluid is injected maintaining the velocity of the cold layer (which is
zero for one-layer case $\epsilon\ll 1$). In the latter case, as the injected
fluid is accelerated to the velocity of the hot layer, it exerts drag on the
hot layer, which is reflected in the friction-like velocity dependence in the
coupling terms in Eqs.~(\ref{sys1})--(\ref{sys1end}).\footnote{Such drag converts
some energy from the flow in the hot layer into heat.  According to our model,
heating represents injection of fluid into the top layer, so, if one is to be
precise, the heating function should be renormalized to account for the
frictional heating.  
When
included in simulations, however, such renormalization accounted for no more
than a $25\%$ increase in front speeds, while not affecting the qualitative
picture. Henceforth we will ignore the frictional heating. }

The two fronts move with different speeds: front A achieves $20$~km~s$^{-1}$,
while front B has a speed of $60$~km~s$^{-1}$.  
The main qualitative
difference in the structure of these fronts is the variation of ageostrophic
velocity $v_x$ with position inside the front. For front A the cross-front
displacement of a fluid element is symmetric, with a fluid particle returning to
the same $x$ coordinate after the front passes, while for front B the
ageostrophic component is asymmetric, skewed towards higher forward
velocity. With stronger cross-front circulation front B achieves a larger speed
of propagation.  The cause of the larger ageostrophic velocity in case B is the
modification of the geostrophic balance brought by the momentum-coupling
drag. The isallobaric wind relationship is now $v_x = - (g/f^2) [{d \over d t}
\nabla h +(Q/h) \nabla h]$. Since $Q/h\sim 1/t_n$ the two terms in this formula
are of the same order of magnitude, and the cross-front wind is enhanced where
the magnitude of the pressure gradient is growing with time (ahead of the peak
heating), and diminished where the pressure gradient is falling. As the heat
transport depends on the cross-front circulation, the momentum-conserving front
should exhibit larger speed of propagation. This is consistent with the results
obtained in Sec.~\ref{secanalyt} where it was shown that, in general, friction
can increase the front speed.

The nature of the temperature overshoot in Fig. \ref{twofronts}a can
 also be attributed to the differences in ageostrophic velocities between the
 two fronts. In front A, the hot fluid flows towards the back of the front as
 much as it does in the forward direction. This flow increases the temperature
 of the fluid at the back of the front beyond the equilibrium value. After the
 helium fuel is depleted the material cools by radiation, and a ``cooling
 tail'' develops because of the time delay in the start of cooling due to
 the finite speed of the front.  
 Momentum conserving fronts also develop cooling tails,
 but because of the 
 larger propagation speed the front moves through a larger distance before the
 fuel depletes enough for cooling to dominate heating (see discussion in
 Sec.~\ref{coolflow}).

We can derive a formal expression for the front propagation speed by
transforming equations (\ref{sys1}-\ref{sys1end}) for the one-layer model to the
frame comoving with the front.  Expressing the ageostrophic component $v_x$ as
a sum of constant front speed $v_f$ and a residual $v_x'$, and assuming that
in the comoving frame the front is not evolving in time we obtain:
\begin{eqnarray}
{\partial \over \partial x} [h v_x']&=& Q, \\
v_x' {\partial \over \partial x} v_x'&=& -g {\partial \over \partial x} h + f v_y
-{Q+\mu_f\over h} (v_x'+v_f), \\
v_x' {\partial \over \partial x} v_y &=& - f(v_x'+v_f)-{Q+\mu_f\over h} v_y.
\end{eqnarray}
As can be seen from Fig. \ref{twofronts}, the geostrophic velocity $v_y$
reaches an extremum inside the front. Denote all values at this point
with an asterisk.  At the extremal point, $v_x'|_*=-v_f-v_y^*
(Q(h^*)+\mu_f)/h^*$, and we can solve the above system of equations for the
speed of the front:
\begin{equation}
v_f=-\Big[{Q(h^*)-h^* {\partial \over \partial x} v_x|_*  \over 
{\partial \over \partial x} h |_*}+{v_y^*\over f}
{Q(h^*)+\mu_f\over h^*} \Big]  \label{frspeed}
\end{equation}
Momentum coupling and viscous friction effects are contained in the second term
on the right hand side of (\ref{frspeed}). Without them the speed of front
propagation is controlled by the effective heating and the pressure gradient at
the peak of the geostrophic velocity. The effective heating $Q(h^*)-h^*
{\partial \over \partial x} v_x|_*$ is the nuclear energy generation minus the
heat carried by the cross-front flow. When the pressure gradient at the peak is
approximated as $h^*/a_R$ we get that the front speed is set by the
characteristic front width divided by the effective burning time, as in
(\ref{frontspeed4}).  
The momentum-coupling drag term is of the same order of magnitude as the first
term in (\ref{frspeed}), and increases the speed of propagation and the width
of the front by a factor of $2-3$.

\subsection{Multilayer dynamics}

\begin{figure}
\centerline{\psfig{file=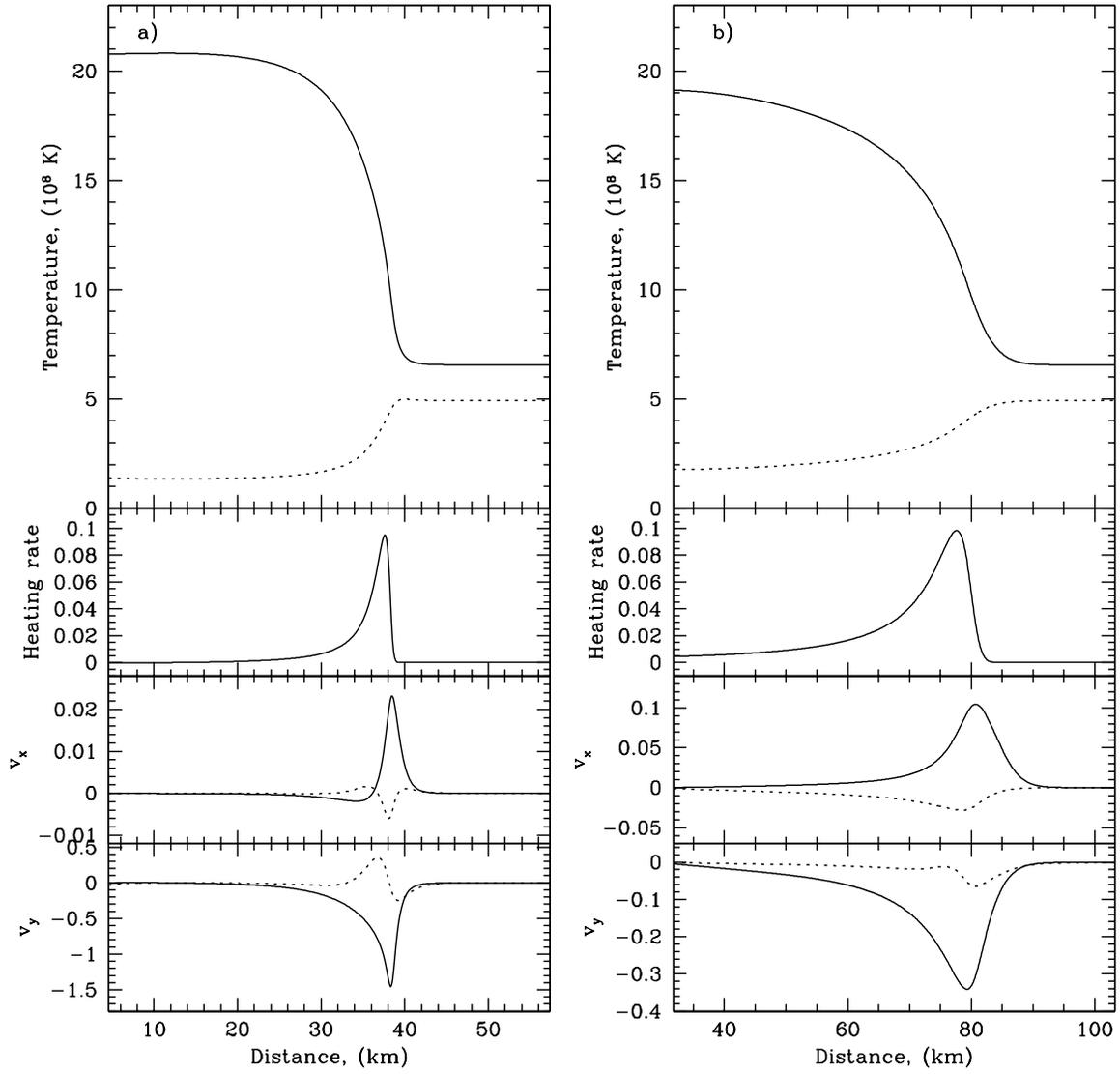,width=6in}}
\caption{ Internal structure of burning fronts in the two-layer model with
$\epsilon=0.2$.  Shown are the thickness (temperature) of the layers with layer
2 (solid) added to layer 1 (dashed), the instantaneous heating rate (in units
of $5\times 10^{19} ~{\rm erg ~g^{-1} s^{-1}}$), cross-front ageostrophic
velocity $v_x$, and tangential geostrophic velocity $v_y$ for two layers
(velocities are in units of gravity wave speed of the cold material $1.5 \times
10^8 {\rm ~cm ~s^{-1}}$).  a) Momentum-conserving front without viscous
friction ($\hat{\mu}_f=0$). For demonstration purposes, $v_x$ and $v_y$ for the
bottom layer are increased by factors of 2 and 5 respectively; b)
Momentum-conserving front with friction ($\hat{\mu}_f=1$). $v_y$ for the bottom
layer is increased by factor of $5$. }
\label{front2l}
\end{figure}

Although illustrative, the one-layer model represents the motion of only the hot
layer of fluid and does not capture all of the dynamics of the front.  In order
to better understand the propagation of the deflagration front through the
neutron star atmosphere we numerically solve the time-dependent system of
equations (\ref{sys1}-\ref{sys1end}) for two layers of isentropic fluid.
Figure \ref{front2l}a demonstrates the structure of the developed
momentum-conserving burning front without interlayer friction propagating from
left to right. In this particular run the density contrast $\epsilon$ was set
to $0.2$, with the 
initial thickness of the lower layer chosen to be three times
the thickness of the top layer.  The first panel shows the evolution of layer
thickness through the front.  As the fluid is pumped to a higher entropy state,
layer 2 expands and layer 1 contracts; the ratio of contraction to expansion is
proportional to the density contrast.  We take the horizontal location of the
peak in the heating rate as the center of the front. From the plot of the
ageostrophic velocities we see that there is a divergence of the fluid at the
top level and a convergence at the bottom level.  The low-entropy fluid is drawn
towards the center of the front and after ignition outflows in the high-entropy
layer. The motion of fluid at the lower layer can also be interpreted as a
response to the gradient in the column depth across the front created by
divergence of the light fluid on top. In addition to creating convergence
towards the center, the overpressure ahead of the front center also forces some
flow to go in the direction of front motion.  This ``snowplow'' effect will
become important in the regime of strong top-bottom coupling; see the next
subsection.

The dominant component of motion in both layers is the geostrophic velocity
(third panel of Fig.~\ref{front2l}a). The direction of this flow is 
different between the
layers: divergence at the top generates anticyclonic motion, while convergence
at the bottom generates cyclonic motion. The fluid velocity in the upper level
is larger than that in the lower level.  The relative velocities in the two
layers depend on the density contrast and on the interlayer friction.

\subsection{Frictional effects}
\label{sec:friction}

We have argued in section \ref{secanalyt} that friction between the top and the
bottom of the atmosphere modifies the geostrophic balance within the front, and
changes the velocity of the flow across the front. The cross-front velocity attains
properties of a diffusive flow governed by the diffusion constant
(\ref{diffcoeff}). Of course, the diffusion analogy is only formal and refers
to the spreading of pressure gradients with time in the presence of friction.
Depending on the strength of frictional coupling, the speed of the front
may be either enhanced or diminished [cf. Eq.~(\ref{vflame22})].  
Recall that in our analytical estimates (\S~\ref{secanalyt}) we
parametrized the strength of friction by the coupling time $t_{\rm
fr}$, while in our two-layer model, described by
eqs. (\ref{sys1}-\ref{sys1end}), the friction strength is parameterized by
$\mu_f/h$, where $\mu_f$ is a constant. To facilitate the comparison
between analytical and numerical results, we define, for our one-layer model,
the characteristic timescale of frictional coupling as $t_{\rm {fr}}=h^*/\mu_f$,
where $h^*$ is the layer height at the peak of the geostrophic velocity inside
the front (in the presence of friction this is approximately but not exactly
the location of peak heating).  In Fig.~\ref{speedfrict} we plot the speed of
the front obtained from one-layer simulations versus the 
dimensionless friction parameter
$\hat{\mu} \equiv (t_{\rm {fr}} f)^{-1}$ and show the fit using
the analytic formula for the front speed in Eq.~(\ref{vflame21}).

 There are several distinct regimes depending on how the timescale of
 frictional forcing compares to the characteristic timescales of the
 problem. When $1/f < t_n < t_{\rm {fr}}$
(or $\hat{\mu}< 0.01$ for our parameters) the effects of friction are
insignificant and the speed is close to the value obtained in section
\ref{structure} after including momentum coupling ($\sim 60 \hbox{~km~s}^{-1}$).  For
$1/f < t_{\rm {fr}} < t_n$ ($0.01 < \hat{\mu} < 1$) the speed increases as
$\sqrt{\hat{\mu}/(1+\hat{\mu}^2)}$ in agreement with (\ref{vflame21}).  The
friction in this regime acts to reduce the geostrophic velocity, which extends
the front over a larger spatial scale. The cross-front diffusion constant
(\ref{diffcoeff}) is maximum when friction acts on the rotation timescale
$\hat{\mu}=1$, and at this value the front speed reaches the peak value ($464
\hbox{~km~s}^{-1}$ for our parameters). The structure of the two-layer model at
$\hat{\mu}=1$ is shown in Fig. \ref{front2l}b. The main distinction of this
case from the case of zero friction (fig. \ref{front2l}a) is the form and the
larger magnitude of the ageostrophic speed, which implies very strong
cross-front circulation.  In the top layer the flow is purely in the forward
direction with a strong return current in the bottom layer. The value of the
geostrophic speed is clearly reduced compared to the no-friction case, and the
strong coupling between the layers causes the flow on the bottom to have the
same anticyclonic direction as the flow in the top layer.

When the friction is further increased ($\hat{\mu}>1$), the cross-front
diffusion (\ref{diffcoeff}) diminishes and the front speed decreases. The
cross-layer coupling now acts to reduce the ageostrophic component of velocity
by making both layers move together. This ``snowplow'' effect eventually stalls
the front when the friction is very large.
Rather than create circulation across the
front, the pressure gradient between the hot and cold material tries to push
forward the whole cold atmosphere, which makes the front steepen up and slow
down.

It is interesting to speculate whether the extraordinarily high front
speeds of hundreds of km~s$^{-1}$ observed in simulations can be expected to
occur in reality.  One does not expect a well-defined burning front to
arise if $t_{\rm {prop}}<t_n$, where $t_{\rm {prop}}\sim\pi R/v_{\rm
  {front}}$ is the timescale on which the formed front would cross the
half-circumference of the neutron star. In other words, it makes no
sense to talk about a well-defined front with a width $\Delta\sim
v_{\rm front} t_n$ that is larger than the size of the star.
This implies that if $v_{\rm {front}}\gtrsim \pi R /t_n \sim 300
\hbox{~km~s}^{-1}$, the front never forms and the star is essentially ignited
simultaneously.  Depending on the physical source of friction in the
neutron star atmosphere there could then be two possible scenarios for the
spread
of nuclear burning.  If the strong friction is temperature dependent, and
increases when the local perturbation grows into runaway (as may be the case
for convective friction), the runaway will begin with no friction present
and,
therefore, will be confined to a hot spot of the size of the Rossby radius. As
the runaway progresses, the friction will increase and the temperature
perturbation will quickly ($\sim 0.1$~s) spread over the neutron star
surface.
If, however, the friction is strong and present all of the time
during the runaway (e.g., if a magnetic field is threading the fuel), an initial hot spot may not appear at all. Rather, the temperature
perturbation
spreads away from the heat source faster than the local thermonuclear
runaway can develop,
and the whole surface is likely to ignite almost simultaneously in a
spherically symmetric fashion.

\section{Global hydrodynamical flows during X-ray bursts and connection to observations}
\label{sectionfour}
\subsection{Likely location of burst ignition} \label{burstloc}

So far we have assumed that the pre-burst conditions are identical
everywhere on the neutron star. However, because the star rotates, the
effective gravity felt by the fluid elements near the equator is
somewhat (up to $\sim 25\%$) smaller than that felt by the fluid
elements near the poles.  As we now argue, this asymmetry implies
that, even if accretion is perfectly spherically symmetric, the fuel
near the equator is likely to reach ignition conditions first.

We assume that, prior to a burst, the accreted material is brought into
corotation with the rest of the star either by effective hydrodynamic
viscosity produced by Rayleigh-Taylor or baroclinic instability
(Fujimoto 1988),  by weak hydromagnetic stresses, or even by
microscopic viscosity (Cumming and Bildsten 2000).  The timescales
for the first two  processes are small compared to the interval between the
bursts, so this assumption is probably accurate as a zeroth-order
approximation.  We also assume that the gas deposited onto the star
can redistribute to achieve hydrostatic balance\footnote{Inogamov and Sunyaev (1999)
have considered the spreading of  material accreted onto a rotating neutron star
from a thin equatorial disk. They have concentrated on the hot radiation-pressure
dominated fluid at the very top of the atmosphere, which has not had time to
frictionally couple to the rest of the star. The nuclear fuel participating
in an X-ray burst is below the hot radiating layer considered by Inogamov and Sunyaev.
We therefore expect that the mid-lattitude ``rings of fire'' found
by Inogamov and Sunyaev will not affect the location of  burst ignition.}.

In hydrostatic equilibrium, the pressure at the bottom of the accreted
ocean should be the same everywhere,
\begin{equation}
p_{\rm bottom}=m(\lambda)g_{\rm eff}(\lambda)={\rm const},
\label{eq4}
\end{equation}
where $m(\lambda)$ is the accreted column density, $\lambda \equiv 
\pi/2 - \theta$ is the latitude, and $g_{\rm eff}=|\nabla
\phi|$ is the effective gravitational acceleration.  Here $\phi$ is
the sum of the gravitational potential, $\phi_{\rm gr}$, and the
centrifugal potential, $\Omega^2 R^2 \cos^2\lambda /2$. The effective
gravity $g_{\rm eff}$ on the equator is less than that on the poles due
to rotation of the star: $\delta g_{\rm eff}/g_{\rm eff}\sim
(\Omega/\omega_K)^2$, where $\omega_K$ is the Keplerian angular
frequency at the neutron-star surface. For a neutron star rotating at
$300-600$~Hz, the relative difference is a few percent (up to $1/4$).
Therefore, the column depth $m(\lambda)$ to a surface of a given pressure is a
factor of $\delta g_{\rm eff}/g_{\rm eff}$ higher on the equator than
on the poles. Now concentrate on two fluid elements, both at a pressure
$p_{\rm bottom}$, one of which is located on the equator while the
other one is somewhere at a latitude $\lambda \neq 0$. Imagine that an
additional amount of gas is accreted, and is allowed to come to
corotation and redistribute itself on the surface of the star in order
to achieve hydrostatic balance. Our fluid elements will now be
compressed to a pressure $p_{\rm bottom}+\Delta p$ (since they are
assumed to be in hydrostatic balance, they must have the same
pressure). But, as evident from Eq.~(\ref{eq4}), the increase in
the column depth, $\Delta m(\lambda)$, must be larger on the equator than away
from it.  Therefore, fuel arriving on the equator will reach a given
column depth faster than the fuel located off-equator.  In other
words, the local accretion rate is inversely proportional to the local
effective gravity strength,
\begin{equation}
\dot{m}(\lambda)=\dot{m}_0{g_0\over g_{\rm eff}(\lambda)},
\label{eq5}
\end{equation}
where $\dot{m}_0=\dot{M}/(4\pi R^2)$ is the average local accretion
rate, and $g_0$ is the average acceleration of gravity, chosen so that
the integral over the surface of the star 
$\int_{-\pi/2}^{\pi/2} (1/2)\cos \lambda [g_0/g(\lambda)]d \lambda=1$.  
The local effective accretion rate seen on the equator is at
least a few percent higher than that seen at the poles.

At a given local accretion rate, which determines the thermal profile
of the accreted ocean, a certan critical column depth is necessary for
the thermal instability and a thermonuclear runaway (see Bildsten 1998
for a review of spherically symmetric ignition conditions). As we
argued above, the critical column depth necessary for a runaway will be
achieved first close to the equator.  We can therefore expect that
ignition happens close to the rotational equator.

\subsection{Burning front propagation on a $\beta$-plane}
\label{betaplanepr}
The speed of the front propagation is larger near the
equator,\footnote{Our equations break down at the equator, because the
horizontal Coriolis force due to vertical motion,
$2\Omega\cos{\lambda} v_z\approx 2\Omega\cos{\lambda}(h/t_n)$,
 is no longer negligible compared to the
Coriolis force due to horizontal motion, $f v_g\approx 2\Omega
\sin{\lambda} (gh)^{1/2}$.  However, this breakdown occurs only for
latitudes $\lambda\leq(h/g)^{1/2}(1/t_n)\sim 10^{-5}$ (for typical
values of $h$ and $t_n$), i.e., very near the equator.}  where
$f\rightarrow0$, than near the poles, where $f\rightarrow f_{max}=2
\Omega$ (cf. Eqs.~[\ref{vfront12}] and [\ref{vflame21}]).  To
understand qualitatively the effect of the latitude dependence of the
Coriolis parameter, we model front propagation on an $x-y$ plane
oriented with $x$ axis east, and $y$ axis -- north.  The Coriolis
parameter is taken to be $f=\beta y$, where $\beta=2\Omega/R$. This
approximates a rotating sphere in the equatorial region, and is known
in geophysical literature as the equatorial $\beta-$plane.

\begin{figure}
\centerline{\psfig{file=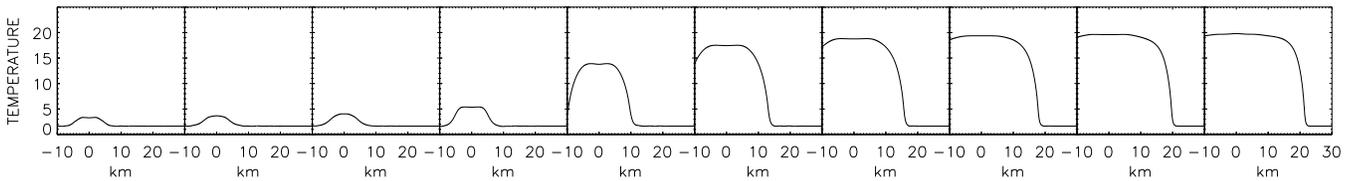,width=7in}}
\caption{ Ignition and propagation on $\beta$-plane. Frames are separated by $0.06$ s}
\label{betaplane}
\end{figure}

\paragraph{$\mathbf1$-D simulations of equatorial ignition.}
We begin in $1$-D with the simulation axis oriented along the $y$ axis
from the equator
to the north pole (and allow variation in quantities only along this
direction).  We solve a one-layer shallow water model without
friction, which is initialized by increasing the temperature on the
equator. A series of snapshots in time is shown in
Fig.~\ref{betaplane}.
Since the Coriolis parameter is zero on the
equator, it is
not possible to establish geostrophic balance there. Gravity waves from
the initial perturbation then propagate towards the pole until they are reflected
from the region with a finite Coriolis parameter at approximately
$y=a_{RE}\equiv({\sqrt{g h}/ \beta})^{1/2}$.  This distance is called the
equatorial Rossby radius and is the characteristic width of the equatorial
waveguide for gravity waves.  The trapped gravity waves are amplified
by thermonuclear energy release on each pass.  For our choice of
initial conditions, after about $t_n \sqrt{g
h}/a_{RE}=t_n (\beta \sqrt{g h})^{1/2} \sim 10^{2-3}$ passes the whole
equatorial waveguide region undergoes a runaway and two geostrophically
supported burning fronts propagate towards the poles as ``walls of fire''.
The propagating fronts steepen and slow down as expected because of
the increasing $f$. For burning on a rotating neutron star the width and speed
of the front decrease by a factor of $2 \Omega/(\beta \sqrt{g
h})^{1/2}\approx 4$ as the front propagates from the equator to the pole.

\paragraph{$\mathbf2$-D model for front propagation.}
From Eq.~(\ref{vflame21}) we can infer the front speed on the $\beta$-plane:
\begin{equation}
v_{\rm flame}(\tilde{x}, \tilde{y})=v_{\rm eq}{1\over\sqrt{1+\tilde{y}^2}},
\label{vbeta}
\end{equation}
where $\tilde{x}=(\beta t_{\rm fr})x$, 
$\tilde{y}=ft_{\rm fr}=(\beta  t_{\rm fr})y$,
and $v_{\rm eq}$ is the speed of the front at the equator. We have 
assumed for simplicity that $t_{\rm fr}\ll t_n$; if this is not the 
case, then in the above expressions $t_{\rm fr}$ must
be replaced by $t_{\rm fr}t_n/(t_n+\eta t_{\rm fr})$.

\begin{figure}
%\unitlength = 0.0011\textwidth
%\begin{center}
%\begin{picture}(275,200)(0,0)
%\put(0,0){\makebox(275,200){ \epsfxsize=275\unitlength \epsfysize=275\unitlength
%\epsffile{burning0.eps}}}
%\put(30,215){\makebox(0,0){\large a)}}
%\put(-1,100){\makebox(0,0){\large $\tilde{y}$}}
%\put(150, -45){\makebox(0,0){\large $\tilde{x}$}}
%\end{picture}
%\hspace{9\unitlength}
%\begin{picture}(275,200)(0,0)
%\put(0,0){\makebox(275,200){\epsfxsize=275\unitlength \epsfysize=275\unitlength
%\epsffile{burning1.eps}}}
%\put(30,215){\makebox(0,0){\large b)}}
%\put(-1,100){\makebox(0,0){\large $\tilde{y}$}}
%\put(150, -45){\makebox(0,0){\large $\tilde{x}$}}
%\end{picture}
%\hspace{9\unitlength}
%\begin{picture}(275,200)(0,0)
%\put(0,0){\makebox(275,200){\epsfxsize=275\unitlength \epsfysize=275\unitlength
%\epsffile{burning2.eps}}}
%\put(30,215){\makebox(0,0){\large c)}}
%\put(-1,100){\makebox(0,0){\large $\tilde{y}$}}
%\put(150, -45){\makebox(0,0){\large ${\tilde{x}}$}}
%\end{picture}
%\end{center}
%\vskip .3in
\centerline{\psfig{file=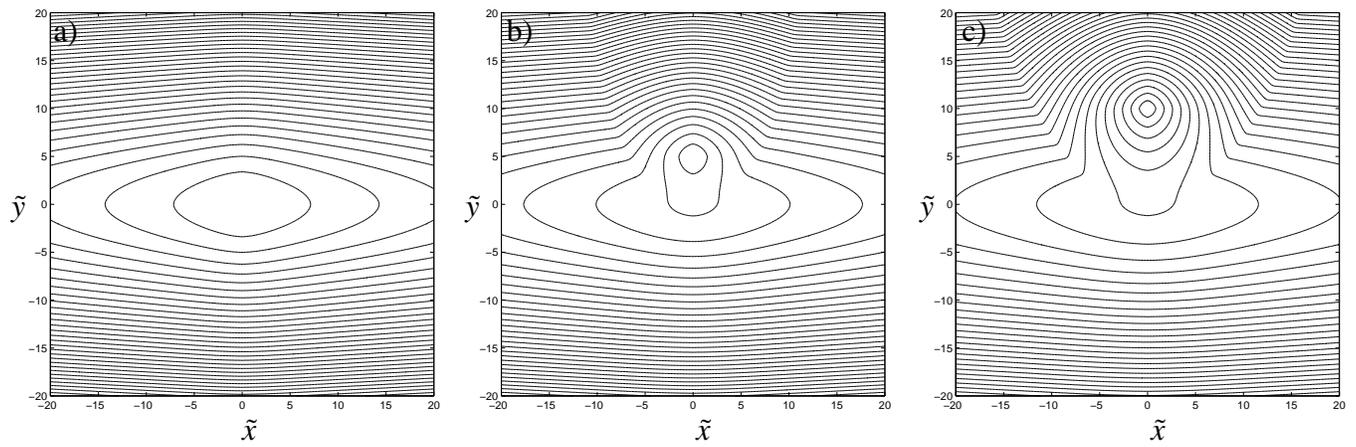,width=7in}}
\caption{Two-dimensional propagation of burning front on $\beta$-plane
for different ignition locations.
Contour lines reflect front position at equal time intervals. a) Equatorial
ignition; b) and c) Ignition at higher latitudes. The dimensionless coordinates
$\tilde{x}$ and $\tilde{y}$ are defined in \S~\ref{betaplanepr} of the text.}
\label{frontprop}
\end{figure}

As was discussed in Section~\ref{locheat}, ignition happens over a patch
of a finite size which is determined by the Coriolis parameter
and the strength of friction in the atmosphere.
However, for the current discussion, we assume that 
ignition is highly localized near some  point on
the $\beta$-plane.  In Fig.~\ref{frontprop} we show how the front 
line develops 
depending on the location of the ignition point, $y_0=f_0 t_{\rm fr}$, where
$f_0$ is the Coriolis parameter at the ignition location.
To compute the evolution of the front line we used 
the Fast Marching Method on a triangulated mesh
(Sethian and Vladimirsky, 2000) with speed function given by (\ref{vbeta}).\footnote{
We thank Alexander Vladimirsky for introducing us to this powerful
method and helping to produce Fig.  \ref{frontprop}.}

Below, we comment separately on the case where the ignition point is
on the equator, and on the case when the atmosphere is ignited at some non-zero
latitude.

\paragraph{\it Equatorial ignition [Fig.~\ref{frontprop}(a)].~} 
The front is roughly a circle
until it reaches $|\tilde{y}|\simeq 1$, i.e.,~until
it reaches the latitude where $f\approx 1/t_{\rm fr}$
and the front speed is significantly
reduced.  Burning then quickly spreads along the equator, and the
front line is nearly parallel to the equator when
the flame reaches $\tilde{y}=ft_{\rm fr}\sim 6$. For a typical strongly
accreting neutron star
in an LMXB the Coriolis parameter at the pole is $4\hbox{--}7\times
10^3$~rad~s$^{-1}$, and the typical frictional coupling strength
$1/t_{\rm fr}$ is anywhere between $~10\hbox{~s}^{-1}$ for
pure momentum coupling,  and
$10^3\hbox{~s}^{-1}$  for frictional coupling by strong
convective turbulence. Our simple $2$-D model then indicates
that  the front becomes parallel to the equator as soon as it reaches 
the  mid-latitudes, or perhaps even closer to the equator.\footnote{When the
front is parallel
to the equator, the flows associated with the front are also parallel
to the equator; such flows are referred to as zonal currents in geophysical
literature.  In this paper we do not consider in detail the 
nonaxisymmetric instabilities
which may be present when the strength of the zonal current is 
latitude-dependent, but see \S \ref{sec:vortices-in-tail}.}

What is the observational significance of this result? The asymmetry
between north-south and east-west front propagation speeds, and the
dependence of the speed on the convective coupling strength may
explain why only a fraction of X-ray bursts shows detectable nearly
coherent oscillations {\it during the burst rise}.  Indeed, during the
burst rise, while the flame, ignited at a spot, is still spreading
around the star, rotational modulation of the X-ray flux should be
evident. Yet it is observed only in some bursts.  A possible
explanation previously proposed in the literature (Miller~2000 and
references therein) is  that the oscillations in the X-ray flux
are washed out when the burst X-rays scatter off a hot cloud which is
thought to surround strongly accreting neutron stars.  This
explanation, however, does not resolve why there are oscillations in
some bursts and no oscillations in others. Muno et.~al.~(2001)
recently concluded that, in fast ($\sim 600$Hz)  rotators (as inferred
from the burst oscillation frequency\footnote{For one of the bursters,
there is evidence that it spins at half of the burst oscillation
frequency (Miller, 1999; see also Strohmayer 2000).}), 
only the strong radius-expansion bursts
show oscillations during the rise, whereas for slower rotators ($\sim
300$Hz) there is no such correlation.

In our model, the equatorial front speed is larger than that at the
poles, and, as argued in \S~\ref{burstloc}, the burst is likely to
ignite at the equator. If the disparity between the front speed at the
equator and the poles is large (i.e., if $2\Omega t_{\rm fr}\gg 1$) then
the entire equatorial belt is likely to ignite in the beginning of the
burst, and the asymmetry in burning which is needed for burst
oscillations may disappear. Increasing the burst strength enhances the
frictional coupling (makes $t_{\rm fr}$ shorter) and makes the burning pattern more
asymmetric, thus making the burst oscillations during the rise more
likely.  This means that equatorially ignited strong bursts are more
likely to be asymmetric than weak bursts.  What about the dichotomy
between fast ($600$~Hz) and ``slow'' ($300$~Hz) rotators? For slow
rotators, the reduction in equatorial gravity due to rotation,
discussed in \S~\ref{burstloc}, is a factor of 4 smaller, so there may
be a significant probability that bursts ignite off the equator, and,
hence, are more asymmetric regardless of burst strength.

\paragraph{\it Nonequatorial ignition [Fig.\ref{frontprop}~(b), (c)].}~
The front is roughly
a circle so long as the distance to the ignition point 
$\sqrt{\Delta\tilde{x}^2+\Delta\tilde{y}^2}$ is less than $1$. Beyond
this distance, the front becomes increasingly deformed. As the edge 
of the flame approaches the equator, the front accelerates, and the
equatorial
belt is quickly ignited. After this, the evolution is similar
to the equatorial ignition case discussed above.

\subsection{Zonal flows and frequency drifts \label{coolflow}}

A salient feature of burst oscillations is their presence in the tails
of bursts, presumably after the entire ocean has been burned, as well
as the increase of the oscillation frequency as the star cools. We
speculate on the possible {\it origin\/} of the inhomogeneity in the
ashes that gives rise to the oscillations in
\S~\ref{sec:vortices-in-tail}; here we discuss the implications of our
global X-ray burst scenario on the {\it frequency\/} of these
oscillations.  In addition, we predict that yet unidentified
oscillations may be present during the burst rise as well. 

The burning front leaves hot ashes in its wake. These hot ashes then cool on
a characteristic timescale $t_{\rm cool}$, defined as $dh_{\rm hot}/dt\sim
h_{\rm hot}/t_{\rm cool}$.  If burning starts in the equatorial region, as
we have argued, then the equator has a headstart in cooling, and in the cooling
wake the equatorial temperature will be the lowest, increasing towards the
poles. This temperature gradient will drive a zonal thermal wind directed
backwards relative to the neutron-star rotation. If an inhomogeneous feature
(e.g., a vortex--see the next subsection) is trapped in this backward zonal
flow, the frequency of the flux modulation due to this feature will
appear lower
than the neutron-star spin frequency. The speed of the backward flow can be
estimated as follows:
\begin{equation}
v_{\rm flow}\sim{g\over f}{\partial h\over \partial x}\sim {g\over f}{h_{\rm hot}\over t_{\rm cool}v_{\rm flame}}.
\label{vflow}
\end{equation}
The drift frequency corresponding to the flow is given by
\begin{equation}
\omega_{\rm drift}={v_{\rm flow}\over R\sin\theta}\sim
{\omega_K^2\over \Omega\sin(2\theta)}{h_{\rm hot}\over
t_{\rm cool}v_{\rm flame}}.
\label{omegadrift1}
\end{equation}
In
the mid-latitude $v_{\rm flame}\sim R/t_{\rm rise}$, where $t_{\rm rise}$ is
the timescale of the burst rise.  The angular speed of the drift in the
mid-latitude is then 
\begin{eqnarray}
{\omega_{\rm drift}\over \Omega}&\sim& \left({\omega_K\over \Omega}\right)^2{t_{\rm rise}\over
t_{\rm cool}}{h_{\rm hot}\over R} \\ \nonumber 
&=&3.6 {h_{\rm hot}\over R}\left({\nu_K\over 2\hbox{kHz}}\right)^2\left({300\hbox{Hz}
\over \nu_s}\right)^2\left({t_{\rm rise}\over 1\hbox{~s}}\right)
\left({10\hbox{~s}\over t_{\rm cool}}\right);
\label{omegadrift2}
\end{eqnarray}
here $\nu_K$ and $\nu_s$ are surface Keplerian and spin frequency, respectively.
For comparison, the frequency due to the radial expansion of the
burning layer (Strohmayer et.~al.~1998, Cumming and Bildsten 2000) is
\begin{equation}
\left({\omega_{\rm drift}\over \Omega}\right)_{\rm lift-up}\sim {h_{\rm hot}\over R}.
\label{CB1}
\end{equation}
The geostrophic drift from Eq.~(\ref{omegadrift2}) and the lift-up
drift from Eq.~(\ref{CB1}) are independent of each other, since the
lift-up drift is related to the {\it horizontal} component of the spin
vector which we have neglected in our calculations. Both drifts must
be present in real bursts. One effect is not obviously dominant over
the other; the ``free'' parameters in Eq.~(\ref{omegadrift2}) decide
which effect is more important for a particular burst.  

Recently
Van Straaten et.~al.~(2001), Galloway et.~al.~(2001), Wijnands et.~al.~(2001),
and Cumming et.~al.~(2001) showed that the lift-up drift is insufficient to
explain the observed magnitude of the frequency drifts in some bursts.
Perhaps the geostrophic wind could provide the missing piece. A
significant amount of modeling of burst light curves is required to
make a more detailed comparison with observations; this is the subject
of our current research.

Note that from Eq.~(\ref{omegadrift1}) one can infer significantly larger
drifts close to the pole and the equator, than the mid-latitude drifts given
by Eq.~(\ref{omegadrift2}).  However, close to the equator the geostrophic
balance is broken, and Eq.~(\ref{omegadrift1}) is no longer valid. 
The origin of divergence of $\omega_{\rm drift}$ at the pole is that we
formally take the distance to the pole to zero, while keeping non-zero $v_{\rm
flow}$. Perhaps, physically one should not consider distances to the pole
closer than the width of the burning front. 

Like the lift-up drift, the geostrophic drift velocity asymptotes to
zero as cooling continues, since the scaleheights at the equator and
the pole equalize. Thus, the final X-ray modulation frequency
asymptotes to the neutron-star spin.  A possible observational
signature that can distinguish the two effects is the time dependence
of the drift frequency. Consider an atmosphere undergoing black-body
cooling in the tail of the burst. The scaleheight of the atmosphere
then decreases with time as $h(t)=h_{{\rm hot}}(1+ t/t_{c})^{-1/3}$,
where $t_c={ c_p \kappa y^2 /(3 a c T_0^3) }$ is the characteristic
cooling time at the maximum expansion when the temperature is
$T_0$. For $T_0=10^9$ K, $y=5.4\times 10^8$~g~cm$^{-2}$, and
$\kappa=0.03 \rm{~cm^{2}~g^{-1}}$, this cooling time is $2$~s. Since
the geostrophic speed depends on the gradient of the scaleheight, the
X-ray oscillation frequency due to geostrophic drift should change
with time as:
\begin{equation}
{\omega(t)\over \Omega}=1-{\omega^0_{\rm{drift}} \over \Omega} {1 \over 
[1+t/t_c]^{4/3}},
\label{eq:geo-cooling-drift}
\end{equation}
where $\omega^0_{\rm{drift}}$ is the drift frequency at the start of
cooling. Here we assumed that the feature that gives rise to the X-ray
oscillations remains at the same latitude for the entire duration of
the cooling tail of the burst and is carried by the geostrophic flow
(however, see \S~\ref{sec:vortices-in-tail} regarding the possibility
of latitudinal drifts which can modify Eq.~[\ref{eq:geo-cooling-drift}]).

On the other hand, the frequency drift due to the radial expansion of
the atmosphere depends on the scaleheight of the column, rather than
its gradient.  
The oscillation frequency
asymptotes to the neutron star rotation 
frequency as 
$1-({\omega^0_{\rm{lift-up}}}/\Omega)[1+t/t_c]^{-1/3}$.
Since it is likely that in real bursts both the lift-up and the
geostrophic drifts are present, the X-ray oscillation signal will have
a time-dependence corresponding to a superposition of the two
effects. If both drifts are of the same order of magnitude, different
time-dependence of the two effects will result in the initial stage of the
frequency evolution being dominated by the geostrophic drift, 
while the late stage will be dominated by the lift-up drift. 
We note that the time scalings given above assume a very simplistic
model of the evolution of the atmospheric scaleheight during
the cooling. Clearly, more detailed
simulations are necessary for robust comparisons with observations.

As demonstrated in Sections ~\ref{secanalyt} and~\ref{numerical}, the
dominant flow during the burst {\it rise} is a geostrophic
current directed parallel to the front line. For the moment, let us
assume that the friction within the front is weak. Then the
geostrophic current is given by $v_g\sim (gh_{\rm
  hot})^{1/2}\sim1.5-3\times10^8$~cm~s$^{-1}$, and is independent of latitude
[this follows from  eqs. (\ref{thermalwind3}) and (\ref{width3})].
If there is a temperature
inhomogeneity entrained in this current (similar to the
inhomogeneities responsible for oscillations in cooling tails of
bursts which were discussed above), we may expect a flux modulation
at a frequency higher than the spin frequency; it will appear
as an upper sideband of the main burst oscillation frequency during the
rise. 
When the front is at a latitude
$\lambda$, the frequency of this sideband is  $v_g/(2\pi
R\cos\lambda)$. As the flame propagates towards higher latitudes
and carries the inhomogeneity along, the frequency chirps up.

The time evolution of this chirp is easy to estimate.  The speed of
the front depends on latitude as $v_f=v_f^{\rm pole}/\sin\lambda$,
where $v_f^{\rm pole}\sim \sqrt{g h_{\rm hot}}/(2 \Omega t_n)$ is the
speed of the front near the pole.
Expressing the front speed as 
 $v_f=R d \lambda / d t$, we 
solve this equation for the front latitude as a function of time: $\cos
\lambda =1- t/t_p$, where $t_p=R/v_f^{\rm pole}$.
The time evolution of the sideband frequency is then
\begin{equation}
\nu_{\rm chirp} ={v_g \over 2 \pi R \cos \lambda} = 
{\nu^{\rm eq}_{\rm chirp} \over 1- t/t_p}.
\end{equation}
Here, $\nu^{eq}_{\rm chirp}=(gh)^{1/2}/2\pi R$ is the sideband
frequency in the region of equatorial belt, which, for typical
parameters in our simulations, ranges from $25$ to $50$ Hz. 
The
final frequency of the chirp depends on how close to the pole the
front stops propagating.  If we take this halting distance to be one Rossby
radius $a_R^{\rm pole}\sim 1$ km, the final frequency of the chirp is
ten times the equatorial value, or $250-500$ Hz.  

When friction is present, the speed of the zonal flow within the
front is
\begin{equation}
v_{\rm geostrophic}\sim {gh_{\rm hot}\over f\Delta}\sim {gh_{\rm hot}\over
ft_n v_{\rm flame}},
\label{vchirp25}
\end{equation}
where $v_{\rm flame}$ is given by Eq.~(\ref{vflame21}).
We have then
\begin{equation}
v_{\rm geostrophic}\sim 
\left({gh_{\rm hot}\over {t_n/t_{\rm fr}}}\right)^{1/2}
{{(f^2+1/t_{\rm fr}^2)^{1/2}}\over
f}.
\label{vchirp26}
\end{equation}
Close to the equator, where the Coriolis parameter $f$ is small compared to
the rate of frictional coupling $1/t_{\rm fr}$, the chirp frequency decreases as 
the front moves away from the equator. However, as soon as the front reaches the
region with $f\sim 1/t_{\rm fr}$, the chirp frequency increases as the front moves 
further towards
the pole. When $f$ is comparable to or greater than $1/t_{\rm fr}$, 
the magnitude of the
chirp frequency is smaller by a factor of $\sim \sqrt{t_n/t_{\rm fr}}$ than 
what it would
be if the friction was absent. If measured, the time-dependence and the 
magnitude 
of the chirp frequency  could be used to discern the importance of the 
effective viscosity during the burst.

\subsection{Vortices and oscillations in the tail}
\label{sec:vortices-in-tail}
So far we have not addressed the nature of the perturbations that lead to
 the oscillations in the X-ray flux in the burst tail.  
The speed of the cooling flow described in the previous subsection is
latitude-dependent [cf.~Eq.~(\ref{vflow})]. Thus we can expect a strong zonal
shear, of the type that is observed in the atmospheres of giant planets. Such
a zonal shear is known to be unstable to formation of vortices; sometimes --- 
like
in the case of Jupiter --- these vortices become large and occupy a significant
fraction of the atmospheric surface. We speculate that these vortices do
 form in a cooling wake of an X-ray burst, and are 
responsible for modulation of the X-ray flux in a burst tail. 
Currently we are constructing a $2$-D shallow water simulation to 
address this scenario. Two-dimensional simulations will also allow us to
study several other important effects which are not contained in the 
$1$-D model. In particular, in the presence of a meridionally
varying Coriolis parameter,  vortices tend to acquire a drift even 
if there is no background flow. This effect is known 
as ``$\beta$-drift'' in the geophysical literature 
(see, e.g. Chan and Williams, 1987),
 and is responsible
for the northwest drift of tropical hurricanes on the Earth. 
The direction of the drift depends on the sense of rotation of the vortex, 
with cyclones drifting northwest and anticyclones southwest. 
In order to estimate the velocity of this drift we use the empirical
formula from Smith, Li and Wang (1997): $v_{\beta-{\rm drift}}\sim r_m 
\sqrt{v_m \beta}$, where $r_m$ is the radius where the maximum fluid 
velocity $v_m$ relative to the vortex
guiding center is reached. For a ``hot spot''
(anticyclonic) vortex with a radius equal to the Rossby radius ($\sim 1$~km), 
and maximum internal 
speed $v_m \sim 10^8$~cm~s$^{-1}$ on a star with $\beta=
4\times 10^{-3}$~cm$^{-1}$~s$^{-1}$, 
the drift speed is of order $600$~km~s$^{-1}$. Depending on the direction 
of drift on the surface of the star this effect can yield an
X-ray oscilation frequency of up to $10$~Hz lower than the spin frequency of 
the star. It remains to be seen how burning inside the vortex and potential
nonaxisymmetric instabilities (e.g. development of spiral arms) modify
this estimate and affect the evolution of vortices in two dimensions. 

\subsection{MHD dynamo and coherence of pulsations}
The burning front may be an ideal environment for an MHD dynamo\footnote{We
thank Maxim Lyutikov for alerting us to this fact.}: it is likely to have both 
turbulence and, at least initially, strong shear; additionally, the 
typical edde overturn time
is much less than $t_n$. The equipartition value of the magnetic field is
\begin{equation}
B_{\rm eq}\sim\sqrt{4\pi\rho}v_c\simeq 3\times 10^9 
\left({\rho\over 10^6\hbox{~g~cm}^{-3}}\right)^{1/2} 
\left({v_c\over 10^6\hbox{~cm~s}^{-1}}\right)\hbox{G},
\label{Bfield}
\end{equation}
where, for concreteness, we have assumed that the turbulence is
of convective origin. The coherence length of the dynamo-generated
magntetic field is of the same order of magnitude as
 the ocean scaleheight.

As discussed in the Introduction, oscillations observed in the tails
of X-ray bursts are highly coherent, with $Q$'s of few thousand over
the duration of the burst (van der Klis 2000). Both lift-up
(Strohmayer 1998; Cumming and Bildsten 2000) and geostrophic models
for the drift have difficulty accounting for this coherence: the
burning ocean is strongly sheared, with the top of the scaleheight
moving at a different speed than the bottom. Cumming and Bildsten
(2000) argued that convection might enforce the vertically rigid
rotation of the burning ocean, although they pointed
out that this is far from certain since convective turnover
time is comparable to the spin period. Numerical models of bursts indicate
that strong convection is indeed present when the ocean is ignited and
rises, but dies quickly when the fuel is exhausted and cooling begins.
It therefore seems unlikely that the coherence of burst oscillations 
in the cooling tail of
the burst could be accounted for by convectively enforced vertically rigid
rotation.

We propose a different idea.  The B-field generated by the burning front
will dynamically couple the top and the bottom of the cooling
ocean\footnote{
It seems reasonable to assume that the small-scale magnetic field,
with coherence length
of $\sim10$~m (the ocean scaleheight) will not have a substantial effect
on the large-scale ($\sim$1~km, roughly the width of the front)
lateral shear.}
 on the
timescale (see, e.g., section $4$ of Cumming and Bildsten 2000)
\begin{eqnarray}
t_{\rm couple}&\sim& {4\pi\rho h v_{\rm flow}\over B^2}\sim 
{hv_{\rm flow}\over v_c^2} \\\nonumber 
&=&0.01{\rm{~s}}
\left({h\over 10^3\hbox{~cm}}\right)\left({v_{\rm flow}\over 3\times 10^7\hbox{~cm~s}^{-1}}
\right)\left({10^6\hbox{~cm~s}^{-1}\over v_c}\right)^2,
\label{tcouple}
\end{eqnarray}
which is typically much shorter than the duration of the
oscillations.  Moreover, the coherence of the oscillations, $Q\sim
t_{\rm cool}/t_{\rm couple}$ is of the order of magnitude of observed
values. We
therefore argue that the vertically rigid rotation necessary for the burst
coherence may be enforced by the small-scale magnetic field generated by the
dynamo in the burning front sweeping through the ocean during the
burst rise. Note that after the burst 
this small-scale magnetic field is confined to the new
ashes,
while the freshly accreted matter can remain unmagnetized.  

\section{Conclusions}
We have constructed a new model for the spreading of deflagration
fronts during type I X-ray bursts. In contrast to previous models, we
take into account the horizontal hydrodynamical flows arising due to
the radial expansion of the burning atmosphere/ocean and the action of
the Coriolis force due to rapid rotation typical for strongly
accreting neutron stars in LMXBs.  Our mechanism of heat transport
 relies on coherent hydrodynamical flows across the
front that are set up as the fluid attains momentum balance, and
achieves speeds of front propagation in excess of tens of kilometers
per second, necessary to account for the observed sub-second rise
times of X-ray bursts.  Previous workers invoked large-scale
convective turbulence (on scales much larger than the vertical
scaleheight) in order to obtain comparable front speeds.
The speed of the front is analytically estimated in
Eq.~(\ref{vfront12}) for the case when convective viscosity is not
important, and in Eq.~(\ref{vflame21}) for the case when the 
layers of the burning ocean are frictionally coupled (by, e.g.,
convection).

In addition to the analytical arguments, we constructed and
numerically evolved a two-layer shallow-water model of the burning
layers.  Our simulations agree very well with our analytical
estimates, and we found that the physics of propagation of
geostrophically supported fronts is independent of the details of the
heating function, as long as it is a single-peaked function of
temperature.  This agreement gives us confidence that our results are
valid for models with more realistic microphysics.

We have outlined the behavior of global hydrodynamical flows in the
neutron-star atmosphere/ocean during the burst, and showed that these flows
may explain many of the features of observed bursts. The very short
rise times of X-ray bursts are easily accounted for if the speed of
the burning front is set by the requirement of geostrophic
balance. The lack of burst oscillations in many bursts may be due to
the fact that the speed of the burning front is very dependent on the
location on the neutron star surface, and favors rapid propagation
along near-equatorial latitude bands that wipes out asymmetry, 
rather than simple
spreading of a hot spot that maintains asymmetry. The rather large
frequency drifts of burst oscillations in tails of some bursts can be
accounted for only if the lift-up drift due to the radial expansion
(Strohmayer 1997, Cumming and Bildsten 2000, Cumming et al.~2001) is combined with the
geostrophic drift due to zonal flows in the cooling wake of the
burning front.  We argue that the burning front will generate strong
($\sim 10^9$~G) small-scale magnetic field, and that this magnetic
field will enforce a vertically rigid flow of the ocean in the wake of
the burning front.  The vertical rigidity of the flow explains the
observed high degree of coherence of the burst oscillations during the
cooling tails of bursts, after convection has subsided and can no
longer account for the coherence.  We conjecture that strong zonal
currents during the burst may lead to the formation of vortices of the
type observed in the atmospheres of giant planets. These vortices may
be responsible for X-ray flux oscillations in the burst tail.  In
addition, we predict the presence of yet unobserved chirps during
rises of bursts. Detection of such chirps (which could have been
missed in previous observations because of the large frequency spans
that they cover) will tell us about the details of burning front
propagation, such as its velocity or the latitudinal extent of the
surface of the star that is covered by nuclear burning.

Our model is incomplete in several respects.  We showed that the
burning front on a rotating neutron star is a site of strong vertical
and horizontal differential shear, and that this shear flow can
transport entropy across the front. However, we omitted the
small-scale mixing and heat diffusion, and thus we have not showed
that this heat can indeed be delivered into the cold fuel, heat it up,
and ignite it as the burning front propagates.  Since flows with
strong shear tend to have both local and global hydrodynamical
instabilities, we are confident that such heat exchange does occur and
is robust.  
Only $2$- or $3$-D hydrodynamical simulations will be able to
authoritatively settle this issue, and we have outlined how our
results would be modified if the mixing exists, but is not as
efficient as we have assumed.  In addition, such simulations will be
useful for understanding the dynamical (frictional) coupling between
the layers of the burning ocean, which, as we showed, has a substantial
effect on the speed of propagation of the burning front. Finally,
further modeling of the global nonaxisymmetric instabilities arising
in the horizontal shear flows may confirm our conjecture of trapped
vortices as the origin of flux modulation at late times during X-ray
bursts.

In this paper we have demonstrated the  usefulness of the
shallow-water model for understanding X-ray bursts. Two-dimensional simulations
utilizing the shallow-water model on a sphere will complement
future vertically-resolved hydrodynamical simulations, and
will allow us to generate realistic
lightcurves and make more refined predictions of frequency behavior of
burst oscillations.

\acknowledgments
First and foremost we would like to thank Lars Bildsten and Andrew Cumming for
getting us interested in the problem, and for their persistent
encouragement. We have benefited greatly from discussions with Phil Arras,
Andy Ingersoll, Maxim Lyutikov, Phillip Marcus, 
Marten van Kerkwijk, Alexander Vladimirsky, and Yanqin Wu.  
We thank Sarah O'Donnell for suggestions which helped to improve
the prose of our manuscript. AS thanks Jonathan Arons
for his endless patience, support, and insightful comments.  The project began
at the ITP, Santa Barbara, during the workshop on Spin and Magnetism of Young
Neutron Stars. YL and GU benefited from visits to Caltech and UC Berkeley,
respectively.  AS and YL are supported by the Theoretical Astrophysics Center
at Berkeley, GU is supported by Lee DuBridge fellowship at Caltech.

\begin{appendix}
\section{Formal derivation of model equations}
The shallow-water equations (\ref{sys1}-\ref{sys1end}) for incompressible fluid
can be derived as a formal limit of the Euler equations for compressible 
fluid. 
This is useful for establishing the connection between the entropy sources in 
compressible fluid and mass sources which are used to simulate heating in
the incompressible case. Using the continuity equation and the ideal gas law, 
the heat equation (\ref{heateq}) can be
rewritten as:
\begin{equation}
(1-\kappa) {d T \over d t} = - \kappa T {\mathbf{\nabla}} \cdot \vec{v} +  {J \over c_p},
\label{heateq1}
\end{equation}
where $\kappa=1-1/\gamma$. In the limit of an incompressible fluid, 
$\gamma\to \infty$ and  $\kappa \to 1$. We can then equate the 
right hand side of (\ref{heateq1}) to zero. 
We consider a layer of uniform temperature $T_0$ 
and height h, which in our model is equal to the scaleheight $c_p T_0/g$.
Integrating (\ref{heateq1}) in the vertical direction over the layer height,
we get:
\begin{equation}
h {\mathbf \nabla_{||}} \cdot \vec{v} + h (v_z|_{z=h} - v_z|_{z=0}) 
= {J\over g}.
\end{equation}
We assume a hard surface at the bottom of the layer
($v_z|_{z=0}=0$), and use $v_z|_{z=h} = d h/dt$ as the velocity of the 
surface. This yields the continuity equation as 
in (\ref{sys1}-\ref{cont2}) with 
effective mass source $Q=J/g$ representing heating. 
The momentum equations (\ref{mom1}-\ref{sys1end}) directly 
follow from the Euler equations for a constant density layer in hydrostatic
balance.
\end{appendix}

\newpage

\end{document}